\documentclass{jfm}

\usepackage{graphicx}
\usepackage{natbib}

\ifCUPmtlplainloaded \else
  \checkfont{eurm10}
  \iffontfound
    \IfFileExists{upmath.sty}
      {\typeout{^^JFound AMS Euler Roman fonts on the system,
                   using the 'upmath' package.^^J}%
       \usepackage{upmath}}
      {\typeout{^^JFound AMS Euler Roman fonts on the seystem, but you
                   dont seem to have the}%
       \typeout{'upmath' package installed. JFM.cls can take advantage
                 of these fonts,^^Jif you use 'upmath' package.^^J}%
      }
  \else
  \fi
\fi

\ifCUPmtlplainloaded \else
  \checkfont{msam10}
  \iffontfound
    \IfFileExists{amssymb.sty}
      {\typeout{^^JFound AMS Symbol fonts on the system, using the
                'amssymb' package.^^J}%
       \usepackage{amssymb}%
       \let\le=\leqslant  
       \let\ge=\geqslant  
      }{}
  \fi
\fi

\ifCUPmtlplainloaded \else
  \IfFileExists{amsbsy.sty}
    {\typeout{^^JFound the 'amsbsy' package on the system, using it.^^J}%
     \usepackage{amsbsy}}
    {}
\fi

\newsavebox{\astrutbox}
\sbox{\astrutbox}{\rule[-5pt]{0pt}{20pt}}

\newcommand\eg{e.g.\ }

\title[Energetics and thermodynamics of turbulent molecular diffusive mixing]
{On the energetics of stratified turbulent mixing, irreversible 
thermodynamics, Boussinesq models, and the ocean heat engine controversy}

\author[R. Tailleux]%
{R\ls \'{E}\ls M\ls I\ns T\ls A\ls I\ls L\ls L\ls E\ls U\ls X$^1$%
  \thanks{Present address: Department of Meteorology,
University of Reading, Earley Gate, PO Box 243, Reading RG6 6BB, 
United Kingdom.}}

\affiliation{$^1$Department of Meteorology,
University of Reading, Earley Gate, PO Box 243, \\ Reading, RG6 6BB, 
United Kingdom}

\pubyear{1996}
\volume{538}
\pagerange{119--126}
\date{22 August 2008 and in revised form ??}

\begin{document}

\maketitle

\begin{abstract}

  In this paper, \cite{Winters1995}'s available potential energy (APE)
framework
is extended to the fully compressible Navier-Stokes equations, with the 
aims of: 1) 
clarifying the nature of the energy conversions taking place in
turbulent thermally-stratified fluids; 2) clarifying the role of surface
buoyancy fluxes in \cite{Munk1998}'s
constraint on the mechanical energy sources of stirring required to 
maintain diapycnal mixing in the oceans. The new framework reveals
that the observed turbulent rate of increase in the background gravitational
potential energy $GPE_r$, commonly thought to occur at the expenses of the
diffusively dissipated $APE$, actually occurs at the expenses of internal
energy, as in the laminar case. The $APE$ dissipated by molecular diffusion,
on the other hand, is found to be converted into $IE$, similarly as the
viscously dissipated kinetic energy $KE$. Turbulent stirring, therefore,
does not introduce a new $APE/GPE_r$ mechanical-to-mechanical energy 
conversion, but
simply enhances the existing
$IE/GPE_r$ conversion rate, in addition to enhancing
the viscous dissipation and entropy production rates. This in 
turn implies that molecular diffusion contributes to the dissipation of
the available mechanical energy $ME=APE+KE$, along with viscous dissipation.
This result has important implications for the interpretation of the 
concepts of mixing efficiency 
$\gamma_{mixing}$ and flux Richardson number $R_f$, for which new
physically-based definitions are proposed and contrasted with previous 
definitions.

\par

  The new framework allows for a more rigorous and
general re-derivation from first principles
of \cite{Munk1998}'s constraint also valid for a
non-Boussinesq ocean:
$$
    G(KE) \approx \frac{1-\xi\, R_f}{\xi \,R_f} W_{r,forcing} = 
      \frac{1+(1-\xi)\gamma_{mixing}}{\xi \,\gamma_{mixing}}W_{r,forcing},
$$
where $G(KE)$ is the work rate done by the mechanical forcing, 
$W_{r,forcing}$ is the rate of loss of $GPE_r$ due to high-latitude cooling,
and $\xi$ a nonlinearity parameter such that $\xi=1$ for a linear equation
of state (as considered by \cite{Munk1998}), but $\xi<1$ otherwise.
The most important result is that $G(APE)$, the work rate done
by the surface buoyancy fluxes, must be numerically as large as 
$W_{r,forcing}$, and therefore as important as the mechanical forcing
in stirring and driving the oceans. As a consequence, the overall mixing
efficiency of the oceans is likely to be larger
than the value $\gamma_{mixing}=0.2$ presently used, thereby 
possibly eliminating the apparent shortfall in mechanical stirring energy
that results from using $\gamma_{mixing}=0.2$ in the above formula.

\end{abstract}

\section{Introduction}

\subsection{Stirring versus mixing in turbulent stratified fluids}

  As is well known, stirring by the velocity field greatly 
enhances the amount of irreversible mixing due to molecular diffusion
in turbulent stratified fluid flows, as compared with the laminar case.
A rigorous proof of this result exists for thermally-driven Boussinesq
fluids for which boundary conditions are either of no-flux or
fixed temperature. In that case, it is possible to show that
\begin{equation}
    \Phi = \frac{\int_V \|\nabla T \|^2 dV}{\int_{V} \|\nabla T_c\|^2 dV},
    \label{Phi_function}
\end{equation}
i.e., the ratio of the entropy production (in the Boussinesq limit)
of the stirred state over that
of the corresponding purely conductive non-stirred state
is always greater than unity, where $T$ and $T_c$ are the temperature
of the stirred and conductive states respectively, the proof being originally
due to \cite{Zeldovich1937}, and re-derived by \cite{Balmforth2003}.
The function $\Phi$ was introduced by \cite{Paparella2002} as a measure
of the strength of the circulation driven by surface buoyancy fluxes.
However, because $\Phi$ is analogous to an average Cox number
(the local turbulent effective diffusivity normalised by the background
diffusivity) \eg \cite{Osborn1972,Gregg1987}, 
it is also representative
of the amount of turbulent diapycnal mixing taking place in the fluid.

\par

 Reversible stirring and irreversible mixing, e.g., see \cite{Eckart1948},
occur in relation
with physically distinct types of forces at work in the fluid. Stirring
works against buoyancy forces by lifting and pulling relatively heavier
and lighter parcels respectively,
thus causing a reversible conversion between kinetic energy $(KE)$ and 
available potential energy $(APE)$. Mixing, on the other hand, is the 
byproduct
of the work done by the generalised thermodynamic forces associated with 
molecular viscous and diffusive processes that
relax the system toward thermodynamic equilibrium, \eg \cite{degroot1962,
Kondepudi1998,Ottinger2005}. 
Thus, stirring enhances the work rate done by the viscous
stress against the velocity field, resulting in enhanced dissipation of
$KE$ into internal energy $(IE)$. Similarly, stirring also enhances 
the thermal entropy production rate associated with the heat transfer
imposed by the second law of thermodynamics, which 
results in a diathermal effective diffusive
heat flux that is increased by the ratio $(A_{turbulent}/A_{laminar})^2$
(another measure of the Cox number),
where $A_{turbulent}$ and $A_{laminar}$ refer to the ``turbulent'' and
``laminar'' area of a given isothermal surface, see \cite{Winters1996} 
and \cite{Nakamura1996}. In the laminar regime, the generalised thermodynamic
forces associated with molecular diffusion are known to cause the conversion
of $IE$ into background gravitational potential energy $(GPE_r)$. From a
thermodynamic viewpoint, it would be natural to expect the stirring to 
enhance the $IE/GPE_r$ conversion, but in fact, the existing literature
usually accounts for the observed turbulent increase in $GPE_r$ as the
result of a ``new'' energy conversion irreversibly converting
$APE$ into $GPE_r$. Clarifying this controversial issue
is a key objective of this paper.

\subsection{The modern approach to the energetics of turbulent mixing}

 The most rigorous existing theoretical framework to understand the
interactions between the different forces at work in a turbulent stratified
fluid is probably the
available potential energy framework introduced by \cite{Winters1995},
for being the only one so far that rigourously separates
reversible effects due to stirring from the irreversible effects due 
to mixing (see also \cite{Tseng2001}).
As originally proposed by \cite{Lorenz1955}, such a framework 
separates the potential energy $PE$ (i.e., the sum of the
gravitational potential energy $GPE$ and internal energy $IE$) 
into its available $(APE=AGPE+AIE)$ and non-available 
$(PE_r=GPE_r+IE_r)$ components, with the IE component being neglected
for a Boussinesq fluid, the case considered by \cite{Winters1995}.
The usefulness of such a decomposition stems from the fact
that the background reference state 
is by construction only affected by diabatic and/or irreversible
processes, so that understanding how the reference state evolves
provides insight into how much mixing takes place in the fluid. 

\par

 In the case of a freely-decaying turbulent Boussinesq stratified
fluid with an equation of state linear in
temperature, referred to as the L-Boussinesq model thereafter, 
\cite{Winters1995} show that the evolution equations for $KE$,
$APE=AGPE$, and $GPE_r$ take the form:
\begin{equation}
     \frac{d(KE)}{dt} = -C(KE,APE) - D(KE),
     \label{KE_equation1}
\end{equation}
\begin{equation}
     \frac{d(APE)}{dt} = C(KE,APE) - D(APE),
     \label{APE_equation1}
\end{equation}
\begin{equation}
     \frac{d(GPE_r)}{dt} = W_{r,mixing} = W_{r,turbulent} + W_{r,laminar},
     \label{GPE_equation1}
\end{equation}
where $C(APE,KE)=-C(KE,APE)$ is the so-called ``buoyancy flux'' measuring
the reversible conversion between $KE$ and $APE$, $D(APE)$ is the diffusive
dissipation of $APE$, which is related to the dissipation of temperature
variance $\chi$, \eg \cite{Holloway1986,Zilitinkevich2008}, 
while $W_{r,mixing}$ is the rate of change in $GPE_r$ 
induced by molecular diffusion, which is commonly decomposed
into a laminar $W_{r,laminar}$ and turbulent $W_{r,turbulent}$ contribution.
All these terms are explicitly defined in Appendix A for the L-Boussinesq
model, as well for a Boussinesq fluid whose thermal expansion increases 
with temperature, called the NL-Boussinesq model. Appendix B further
generalises the corresponding expressions for the fully compressible
Navier-Stokes equations (CNSE thereafter) with an arbitrary nonlinear 
equation of state (depending on pressure and temperature only though).

\par 

  Of particular interest in turbulent mixing studies is the behaviour
of $W_{r,turbulent}$ --- the turbulent rate of increase in $GPE_r$ ---
which so far has been mostly discussed in the context of the L-Boussinesq
model, for which an important result is:
\begin{equation}
           W_{r,turbulent} = D(APE),
           \label{wr_turbulent}
\end{equation}
which states the equality between the $APE$ dissipation rate and
$W_{r,turbulent}$. This result is important, because from the known properties
of $D(APE)$,
it makes it clear that enhanced diapycnal mixing rates fundamentally
require: 1) finite values of $APE$, since $D(APE)=0$ when $APE=0$;
2) an $APE$ cascade transferring the spectral energy of the 
temperature (density) field to the small scales at which molecular diffusion
is the most efficient at smoothing out temperature gradients.
The discussion of the $APE$ cascade, which is closely related to that
of the temperature variance, has an extensive literature related
to explaining the $k^{-3}$
spectra in the so-called buoyancy subrange, both in the atmosphere,
\eg \cite{Lindborg2006} and in the oceans, \eg \cite{Holloway1986,Bouruet1996}.
 Note that because $APE$ is a globally defined
scalar quantity, speaking of $APE$ cascades requires the introduction of
the so-called $APE$ density, noted $\Phi_a({\bf x},t)$ here, for which 
a spectral description is possible, \eg \cite{Holliday1981,Roullet2009,
Molemaker2009}.

\par

\begin{center}
\begin{figure}
\includegraphics[width=13cm]{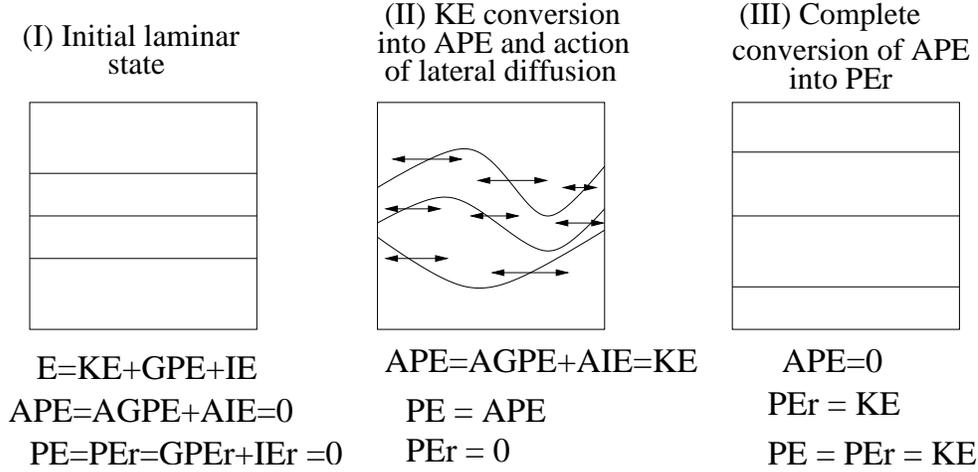}
\caption{Idealised depiction of the diffusive route for kinetic energy
dissipation. (I) represents the laminar state possessing initially no
AGPE and AIE, but some amount of KE. (II) represents the state obtained
by the reversible adiabatic conversion of some kinetic energy into APE, 
which increases APE but leaves the background $GPE_r$ and $IE_r$ 
unchanged; (III) represents the state obtained by letting the horizontal
part of molecular diffusion smooth out the isothermal surfaces until
all the APE in (II) has been converted into background $PE_r=GPE_r+IE_r$.}
\label{diffusiveroute}
\end{figure}
\end{center}

  Eqs. (\ref{KE_equation1}-\ref{GPE_equation1}) exhibit only one type
of reversible conversion, namely the ``buoyancy flux'' associated with 
the $APE/KE$ conversion, and three irreversible conversions, viz.,  
 $D(KE)$, $D(APE)$, and $W_{r,mixing}$, the first one
caused by molecular viscous processes, and the latter two
caused by molecular diffusive processes. A primary goal of turbulence
theory is to understand how the reversible $C(APE,KE)$ conversion and
irreversible $D(KE)$, $D(APE)$, $W_{r,mixing}$ are all inter-related.
In this paper, the focus will be on turbulent diffusive mixing, for the
understanding of viscous dissipation constitutes somehow a separate issue
with its own problems, \eg \cite{Gregg1987}.
The nature of these links is usually explored by estimating the energy
budget of a turbulent mixing event, defined here as a period of intense
mixing preceded and followed by laminar conditions, for which there is
a huge literature of observational, theoretical, and numerical studies.
Integrating the above energy equations over the duration of the turbulent
mixing event thus yields:
\begin{equation}
     \Delta KE = -\overline{C(KE,APE)} - \overline{D(KE)} ,
      \label{KE_budget}
\end{equation}
\begin{equation}
     \Delta APE = \overline{C(KE,APE)} - \overline{D(APE)} ,
    \label{APE_budget}
\end{equation}
\begin{equation}
      \Delta GPE_r = \overline{W}_{r,mixing} = 
      \overline{W}_{r,turbulent} + \overline{W}_{r,laminar} ,
    \label{GPEr_budget}
\end{equation}
where $\Delta (.)$ and the overbar denote respectively the net
variation and the time-integral of a quantity over the mixing event.
Summing the $KE$ and $APE$ equation yields the important 
``available'' mechanical energy equation:
\begin{equation}
     \Delta KE + \Delta APE = -[\overline{D(KE)}+\overline{D(APE)}]<0,
     \label{ME_budget}
\end{equation}
which states that the total ``available'' mechanical energy $ME=KE+APE$ 
undergoes a net decrease over the mixing 
event as the result of the viscous and diffusive dissipation of $KE$
and $APE$ respectively. A schematic of the $APE$ dissipation process,
which provides a diffusive route to $KE$ dissipation,
is illustrated in Fig. \ref{diffusiveroute}.

\subsection{Measures of mixing efficiency in turbulent stratified fluids}

   Eq. (\ref{ME_budget}) makes it clear that turbulent diapycnal mixing
(through $D(APE)$) participates in the total dissipation of available
mechanical energy $ME = KE+APE$. Since $D(APE)$ is non-zero only if
$APE$ is non-zero, turbulent diapycnal mixing therefore requires having
as much of $ME$ in the form of $APE$ as possible. The classical concept of
``mixing efficiency'', reviewed below, seeks to provide a number 
quantifying the ability of a particular turbulent mixing event in 
dissipating $ME=KE+APE$ preferentially diffusively rather than viscously.
From a theoretical viewpoint, it is useful to separate turbulent mixing
events into two main archetypal categories, corresponding to the two cases
where $ME$ is initially
entirely either in $KE$ or $APE$ form. These two cases are treated 
separately, before providing a synthesis addressing the general case.

\par

  At a fundamental level, quantifying the mixing efficiency of a turbulent
mixing event requires two numbers, one to measure how much of $ME$ is 
viscously dissipated, the other to measure how much of $ME$ is dissipated by
turbulent mixing. While everybody seems to agree that $D(KE)$ is the natural
measure of viscous dissipation, it is the buoyancy flux $\overline{C(APE,KE)}$,
rather than $D(APE)$, that has been historically thought to be the relevant
measure of how much of $ME$ is dissipated by turbulent mixing, since it is
the term in Eq. (\ref{KE_budget}) that seems to be removing $KE$ along with
viscous dissipation.
  For mechanically-driven turbulent mixing events, defined here as being
such that $\Delta APE =0$ and $\Delta ME = \Delta KE$, the efficiency of
mixing has been classically quantified by means of two important numbers.
The first one is the so-called flux Richardson number $R_f$,
defined by Linden (1979) as:
 ``the fraction of the change in available kinetic energy
which appears as the potential energy of the stratification'', 
mathematically defined as:
\begin{equation}
     R_f = \frac{\overline{C(KE,APE)}}{|\Delta KE|} = 
     \frac{\overline{C(KE,APE)}}{\overline{C(KE,APE)} + \overline{D(KE)}},
\end{equation}
e.g., Osborn (1980), and the second one is the so-called ``mixing efficiency'':
\begin{equation}
    \gamma_{mixing} = \frac{R_f}{1-R_f} = 
    \frac{\overline{C(KE,APE)}}{\overline{D(KE)}} .
\end{equation}
 It is now recognised, however, that the buoyancy flux represents only an
indirect measure of irreversible mixing, since it physically represents a
reversible conversion between $KE$ and $APE$, while furthermore appearing to
be difficult to interpret empirically,
\eg see \cite{Barry2001} and references therein. 
Recognising this
difficulty, \cite{Caulfield2000} and \cite{Staquet2000} effectively suggested 
to replace $C(\overline{KE,APE})$ by a more direct measure of irreversible
mixing in the above definitions of $R_f$ and $\gamma_{mixing}$. Since 
turbulent diapycnal mixing is often diagnosed empirically from measuring
the net changes in $GPE_r$ over a mixing event, \eg 
\cite{McEwan1983a,McEwan1983b,Barry2001,Dalziel2008}, 
a natural choice is to use
$\overline{W}_{r,turbulent}$ as a direct measure of irreversible mixing,
which leads to:
\begin{equation}
     R_f^{GPE_r} = \frac{\overline{W}_{r,turbulent}}{\overline{W}_{r,turbulent}
     + \overline{D(KE)}}
     \label{Rf_gper}
\end{equation}
\begin{equation}
     \gamma_{mixing}^{GPE_r} = \frac{R_f^{GPE_r}}{1-R_{f}^{GPE_r}} = 
     \frac{\overline{W}_{r,turbulent}}{\overline{D(KE)}} .
     \label{gamma_gper}
\end{equation}
From a theoretical viewpoint, these definitions are justified from the fact
that in the L-Boussinesq model, the following equalities hold:
\begin{equation}
      \overline{C(APE,KE)} = \overline{D(APE)} = \overline{W}_{r,turbulent},
      \label{3_equalities}
\end{equation}
as follows from Eqs. (\ref{KE_budget}) and (\ref{APE_budget}), combined
with Eq. (\ref{wr_turbulent}), when $\Delta APE=0$.
The modified flux Richardson number $R_f^{GPE_r}$ coincides --- for
a suitably defined time interval --- with the cumulative mixing efficiency
$\mathcal{E}_c$ introduced by Caulfield and Peltier (2000), as well as
with the generalised flux Richardson number $R_b$ defined by Staquet (2000),
in which our $\gamma_{mixing}^{GPE_r}$
is also denoted by $\gamma_b$.

\par

 Although Eqs. (\ref{Rf_gper}) and (\ref{gamma_gper}) are consistent with
the traditional buoyancy-flux-based definitions of $R_f$ and $\gamma_{mixing}$
in the context of the L-Boussinesq model, such definitions overlook the
fact that Eq. (\ref{3_equalities}) is not valid in the more general context of
the fully compressible Navier-Stokes equations, for which the ratio
\begin{equation}
   \xi = \frac{W_{r,turbulent}}{D(APE)}
   \label{xi_parameter}
\end{equation}
is in general
less than one, and even sometimes negative,
for water or seawater. For this reason, it appears that
$R_f$ and $\gamma_{mixing}$ should in fact be defined in terms of $D(APE)$,
not $W_{r,turbulent}$, viz., 
\begin{equation}
     R_f^{DAPE} = \frac{\overline{D(APE)}}{\overline{D(APE)}+
    \overline{D(KE)}} ,
    \label{new_rf}
\end{equation}
\begin{equation}
     \gamma_{mixing}^{DAPE} = \frac{\overline{D(APE)}}{\overline{D(KE)}} ,
    \label{new_gamma}
\end{equation}
which we call the dissipation flux Richardson number, and dissipation 
mixing efficiency respectively, to distinguish them from their predecessors.
In our opinion, $R_f^{DAPE}$ and $\gamma_{mixing}^{DAPE}$ as defined by
Eqs. (\ref{new_rf}) and (\ref{new_gamma}) are really the ones that are 
truly consistent with the properties assumed to be attached to those numbers.
Most notably, Eq. (\ref{new_rf}) is the only way to define a flux Richardson
number that is guaranteed to lie within the interval $[0,1]$, since neither
$\overline{C(KE,APE)}$ nor $\overline{W}_{r,turbulent}$ can be ascertained 
to be positive under all circumstances. Since 
Eqs. (\ref{Rf_gper}) and (\ref{gamma_gper}) are still likely to be used
in the future owing to their practical interest, it is useful to provide
conversion rules between the $GPE_r$ and $D(APE)$-based definitions of 
$R_f$ and $\gamma_{mixing}$, viz.,
\begin{equation}
     \gamma_{mixing}^{GPE_r} = \xi \gamma_{mixing}^{DAPE}, 
  \qquad 
     R_{f}^{GPE_r} = \frac{\xi R_f^{DAPE}}{1-(1-\xi)R_f^{DAPE}} .
\end{equation}
These formula require knowledge of the nonlinearity parameter $\xi$,
which measures the importance of nonlinear effects associated with the
equation of state, see \cite{Tailleux2009a} for details about this.
The often-cited canonical value for
mechanically-driven turbulent mixing is $\gamma_{mixing} \approx 0.2$,
which appears to date back from \cite{Osborn1980}, \eg \cite{Peltier2003}.

\par

 The second case of interest, namely buoyancy-driven turbulent mixing,
is defined here as being such that $\Delta KE = 0$ and $\Delta ME = 
\Delta APE$, as occurs in relation with the so-called Rayleigh-Taylor
instability for instance. Eqs. (\ref{KE_budget}) and (\ref{APE_budget})
lead to:
\begin{equation}
       \overline{C(KE,APE)} = -\overline{D(KE)} < 0
       \label{wb_rt}
\end{equation}
\begin{equation}
       \overline{D(APE)} = \overline{C(KE,APE)} - \Delta APE 
     = | \Delta APE | - |\overline{C(KE,APE)}| .
\end{equation}
Eq. (\ref{wb_rt}) reveals that the buoyancy flux is negative this time,
and that it represents the fraction of $ME$ that is lost to {\em viscous}
dissipation, {\em not} diffusive dissipation.
This establishes, if needed, that the buoyancy flux should not be 
systematically interpreted as a measure of irreversible diffusive mixing.
Since 
\cite{Linden1979}'s above definition for the flux Richardson number does not
really make sense for Rayleigh-Taylor instability, an alternative definition
is called for. The most natural definition, in our opinion, is as the 
fraction of $ME$ dissipated by irreversible diffusive mixing, viz.,
\begin{equation}
        R_f = \frac{-\Delta APE + \overline{C(KE,APE)}}{- \Delta APE}
     = 1 - \frac{\overline{|C(KE,APE)|}}{|\Delta APE|},
\end{equation}
which according to Eqs. (\ref{KE_budget}) and (\ref{APE_budget}), 
is equivalent to:
\begin{equation}
        R_f = \frac{\overline{D(APE)}}{\overline{D(APE)} + \overline{D(KE)}},
\end{equation}
with the corresponding value of $\gamma_{mixing}$:
\begin{equation}
 \gamma_{mixing} = \frac{R_f}{1-R_f} = 
      \frac{\overline{D(APE)}}{\overline{D(KE)}} ,
\end{equation}
which are identical to Eqs. (\ref{new_rf}) and (\ref{new_gamma}).
The above results make it possible, therefore, to use $R_f^{DAPE}$ and
$\gamma_{mixing}^{DAPE}$ as definitions for the flux Richardson number
and mixing efficiency that make sense for all possible types of turbulent
mixing events. 

\par

 At this point, a note about terminology seems to be warranted, since
in the case of the Rayleigh-Taylor instability, it is $R_f$ that is 
referred to as the mixing efficiency by some authors, \eg
\cite{Linden1991,Dalziel2008}, rather than $\gamma_{mixing}$. 
Physically, this seems more logical, since $R_f$ is always comprised 
within the interval $[0,1]$, whereas $\gamma_{mixing}$ is not.
Interestingly, \cite{Oakey1982} appears to be the first to define 
$\gamma_{mixing}$ as a: ``mixing coefficient representing the ratio
of potential energy to kinetic energy dissipation''. For this reason,
it would seem more appropriate and less ambiguous to refer to 
$\gamma_{mixing}$ as the ``dissipations ratio''. Unfortunately, it is
not always clear in the literature which quantity the widely used term
``mixing efficiency'' refers to, as it has been used so far to refer to
both $R_f$ and $\gamma_{mixing}$. In order to avoid ambiguities, the
remaining part of the paper only make use of 
the quantities $R_f^{DAPE}$ and $\gamma_{mixing}^{DAPE}$, which for 
simplicity are denoted $R_f$ and $\gamma_{mixing}$ respectively.

 \par
  
  As a side note, it seems important to point out that Rayleigh-Taylor 
instability has the peculiar property that $\Delta GPE_{r,max}$,
the maximum possible increase
in $GPE_r$ achieved for the fully homogenised state, is only 
{\em half} the initial amount of $APE$,  
\eg \cite{Linden1991,Dalziel2008} (at least when $\xi=1$, i.e.,
in the context of the L-Boussinesq model). Physically, it means that less
than 50 \% of the initial $APE$ can actually contribute to turbulent 
diapycnal mixing, and hence that at least 50 \% of it must be eventually
viscously dissipated. As a result, one has the following constraints:
\begin{equation}
    R_f = \frac{\overline{D(APE)}}{|\Delta APE|} 
    = \frac{\xi \overline{W}_{r,turbulent}}{|\Delta APE|} 
   \le \frac{1}{2}
\end{equation}
\begin{equation}
    \gamma_{mixing} \le \frac{\xi/2}{1-\xi/2} \le 1.
\end{equation}
Experimentally, \cite{Linden1991} 
reported values of $R_f \approx 0.3$ 
($\gamma_{mixing}=3/7\approx 0.43$), while \cite{Dalziel2008}
reported experiments in which the maximum possible value $R_f\approx 0.5$
($\gamma_{mixing}\approx 1$) was reached. Owing to the peculiarity of
the Rayleigh-Taylor instability, however, one should refrain from concluding
that $\gamma_{mixing}=1$ or $R_f=0.5$ represent the maximum possible values
for $\gamma_{mixing}$ and $R_f$ in turbulent stratified fluids.
To reach definite and general conclusions about $\gamma_{mixing}$ and $R_f$, 
more general examples of 
buoyancy-driven turbulent mixing events should be studied.
It would be of interest, for 
instance, to study the mixing efficiency of a modified Rayleigh-Taylor
instability such that the unstable stratification occupies only half or less
of the spatial domain, so that $\Delta GPE_{r,max} \ge |\Delta APE|$. In
this case, all of the initial $APE$ could in principle be dissipated by
molecular diffusion, which would correspond to the limits  $R_f=1$ and
$\gamma_{mixing} = +\infty$. Of course, such limits cannot be reached, as
it is impossible to prevent part of the $APE$ to be converted into $KE$,
part of which will necessarily be dissipated viscously, but they are
nevertheless important in suggesting that values of $\gamma_{mixing}>1$ can
in principle be reached, which sets an interesting goal for future research.

\subsection{On the nature of $D(APE)$ and $W_{r,turbulent}$}

\begin{figure}
\center
\includegraphics[width=11cm]{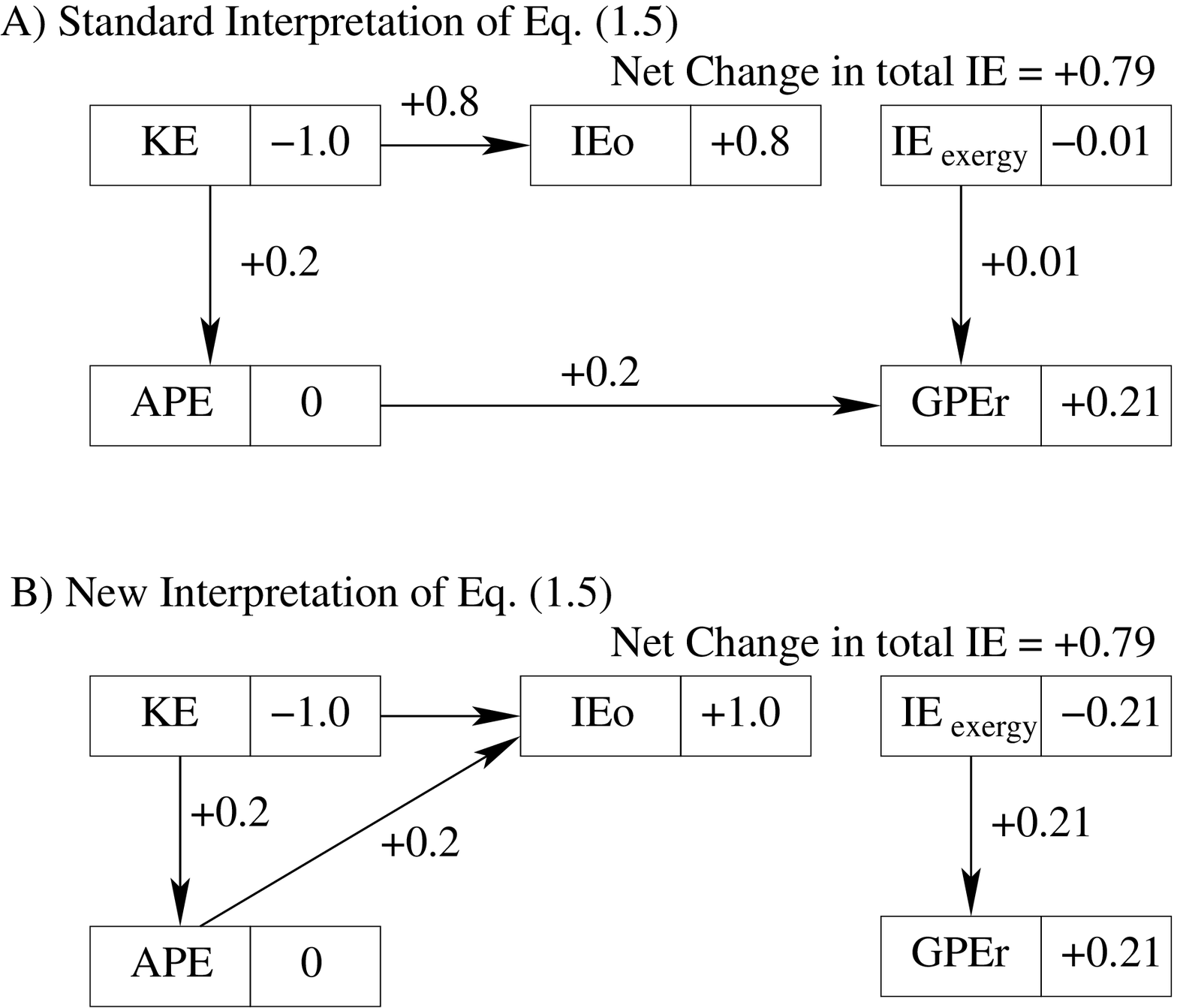}
\caption{(A) Predicted energy changes for an hypothetical turbulent
mixing event under the assumption that the diffusively dissipated $APE$
is irreversibly converted into $GPE_r$; (B) Same as in A under the
assumption that the diffusively dissipated $APE$ is irreversibly 
converted into $IE_0$, as the viscously dissipated $KE$. 
In both cases, the net energy changes in $KE$, $GPE_r$, $APE$, and
$IE$ are the same. The only predicted differences concern the 
subcomponents of the internal energy $IE_0$ and $IE_{exergy}$.}
\label{new_interpretation}
\end{figure}

  Of fundamental importance to understand the physics of turbulent diapycnal
mixing are the nature and type of the energy conversions associated with 
$D(APE)$ and $W_{r,turbulent}$. So far, it seems fair to say that these two
energy conversions 
have been regarded as essentially being one and the same, 
based on the exact equality
$W_{r,turbulent}=D(APE)$ occurring in the L-Boussinesq model, suggesting
that molecular diffusion irreversibly converts $APE$ into $GPE_r$,  
\eg \cite{Winters1995}. Such an interpretation appears to be now
widely accepted, \eg
\cite{Caulfield2000}, \cite{Peltier2003}, \cite{Munk1998}, \cite{Huang2004},
\cite{Thorpe2005} among many others. The main characteristic of this view,
schematically illustrated in the top panel of Fig. \ref{new_interpretation},
is to disregard the possibility that the turbulent increase of $GPE_r$ might
be due the enhancement of the $IE/GPE_r$ conversion rate by the stirring.
In other words, the current view assumes that the work involved in the 
turbulent increase of $GPE_r$ is done by the stirring against buoyancy forces,
not by the generalised thermodynamic forces responsible for entropy production
and the $IE/GPE_r$ conversion. At the same time, the current view seems to
accept that stirring enhances entropy production. But from classical 
thermodynamics, this seems possible only if the work rate done by the 
generalised thermodynamic forces is also enhanced, which in turn should imply
an enhanced $IE/GPE_r$ conversion. 
 
\par

  In order to determine whether the turbulent increase of $GPE_r$ could
be accounted for by a stirring-enhanced $IE/GPE_r$ conversion rate, rather
than by the irreversible conversion of $APE$ into $GPE_r$, it seems useful to
point out that the validity of \cite{Winters1995}'s interpretation seems
to rely crucially on $D(APE)$ and $W_{r,turbulent}$ being
{\em exactly} identical, not only mathematically (as is the case in the
L-Boussinesq model) but also physically. Here, two quantities
are defined as being physically equal if they remain mathematically equal
in more accurate models of fluid flows --- closer to physical ``truth''
in some sense --- such as CNSE
for instance. Indeed, only a physical equality
can define a physically valid energy conversion, as we hope the reader
will agree. As shown in Appendix B, however, which extends \cite{Winters1995}
results to the CNSE, the equality
$D(APE)=W_{r,turbulent}$ is found to be a serendipitous feature of 
the L-Boussinesq model, which at best is only a good approximation, 
the general result being that the ratio
\begin{equation}
     \xi = \frac{W_{r,turbulent}}{D(APE)}.
     \label{xi_ratio}
\end{equation}
usually
lies within the interval $-\infty < \xi <1$ for water or seawater,
and that it 
strongly depends on the nonlinear character of the equation of state.
Whether there exists fluids allowing for $\xi>1$ is not known yet.
An important result is that it appears to be
perfectly possible for $GPE_r$ to {\em decrease} as
the result of turbulent mixing, in contrast to what is often stated in
the literature. This case, which corresponds to $\xi <0$, was in fact
previously identified and discussed by the late Nick Fofonoff in a series
of little known papers, see
\cite{Fofonoff1962,Fofonoff1998,Fofonoff2001}. For this reason, 
the case $\xi < 0$ shall be subsequently referred to 
as the {\em Fofonoff regime}, while the more commonly studied case for
which $W_{r,turbulent}>0$ shall be referred to as the {\em classical regime}.

\par

 The lack of physical equality between $D(APE)$ and $W_{r,turbulent}$
makes \cite{Winters1995}'s interpretation very unlikely, and
gives strong credence to the idea that $W_{r,turbulent}$ actually correspond
to a stirring-enhanced $IE/GPE_r$ conversion rate. If so, what about 
$D(APE)$? In order to shed light on the issue of $APE$ dissipation, it is
useful to recall some well known properties of thermodynamic transformations
associated with the following problem: Assuming that the potential energy
$PE=GPE+IE$ of a stratified fluid increases by $\Delta E$, how is $\Delta E$
split between $\Delta GPE$ and $\Delta IE$? Here, standard thermodynamics
tells us that the answer 
depends on whether $\Delta E$ is added reversibly or irreversibly
to $PE$. Thus, if $\Delta E$ is added reversibly to $PE$ 
(i.e., without entropy change, and for a nearly incompressible fluid), then:
\begin{equation}
       \frac{\Delta GPE}{\Delta E} \approx 1, \qquad
       \frac{\Delta IE}{\Delta E} \ll 1
       \label{adiabatic_transfo}
\end{equation}
while if $\Delta E$ is added irreversibly (i.e., with an increase in
entropy), then:
\begin{equation}
       \frac{\Delta GPE}{\Delta E} \ll 1, \qquad
       \frac{\Delta IE}{\Delta E} \approx 1 ,
       \label{diabatic_transfo}
\end{equation}
i.e., the opposite.
These results, therefore, suggest that when molecular diffusion converts
$APE$ into $PE_r$, the dissipated $APE$ must nearly entirely go into $IE_r$,
not $GPE_r$, in contrast to what is usually assumed 
(The demonstration of Eqs. 
(\ref{adiabatic_transfo}) and (\ref{diabatic_transfo}) is omitted for
brevity, but this follows from the results of Appendix B.)
It follows that what the equality $D(APE)=W_{r,turbulent}$ of the L-Boussinesq
actually states is the equality of the $APE/IE$ and $IE/GPE_r$ conversion 
rates (or more generally, for real fluids, the correlation between the two
rates), not that $D(APE)$ and $W_{r,turbulent}$ are
of the same type. Physically, the two conversion rates 
$W_{r,turbulent}$ and $D(APE)$ appear to be fundamentally correlated
because they are both controlled by molecular diffusion and
the spectral distribution of $APE$, as will be made clear later in the text.

\subsection{Internal Energy or Internal Energies?}

  In the new interpretation proposed above, internal energy is destroyed
by the $IE/GPE_r$ conversion at the turbulent rate $W_{r,turbulent}$, while
being created by the $APE$ conversion at the turbulent rate $D(APE)$. 
Could it be possible, therefore, for the dissipated $APE$ to be eventually
converted into $GPE_r$, not by the direct $APE/GPE_r$ conversion route 
proposed by \cite{Winters1995}, as this was ruled out by thermodynamic
considerations, but indirectly by transiting through the $IE$ reservoir?

\par

  As shown in Appendix B, the answer to the above question is found to be
negative, because it turns out that 
the kind of $IE$ which $APE$ is dissipated into 
appears to be different from the kind of $IE$ being converted into $GPE_r$.
Specifically, Appendix B shows that $IE$ is
indeed best regarded as the sum of distinct sub-reservoirs. In this paper,
three such sub-reservoirs are introduced: the available internal energy
$(AIE)$, the exergy $(IE_{exergy})$, and the dead internal energy $(IE_0)$.
Physically, this decomposition parallels the 
following temperature decomposition:
$T(x,y,z,t)= T'(x,y,z,t) + T_r(z,t) + T_0(t)$, where $T_0(t)$ is the 
equivalent thermodynamic equilibrium temperature of the system, 
$T_r(z,t)$ is Lorenz's reference vertical temperature profile, and
$T'(x,y,z,t)$ the residual. Physically, $AIE$ is the 
internal energy component of \cite{Lorenz1955}'s $APE$, while $IE_0$
and $IE_{exergy}$ are the internal energy associated with the equivalent
thermodynamic equilibrium temperature $T_0$ and vertical temperature
stratification $T_r$ respectively. The idea behind this decomposition can
be traced back to \cite{Gibbs1878}, the concept of exergy being common
in the thermodynamic engineering literature, \eg \cite{Bejan1997}. See 
also \cite{Marquet1991} for an application of exergy in the context of
atmospheric available energetics. A full review of existing ideas related
to the present ones is beyond the scope of this paper, as the engineering
literature about available energetics and exergy is considerable.
  The way it works is encompassed in the following equations:
\begin{equation}
    \frac{d(KE)}{dt} = - C(KE,APE) - D(KE) ,
     \label{ke_new_equation}
\end{equation}
\begin{equation}
    \frac{d(APE)}{dt} = C(KE,APE) - D(APE) ,
    \label{ape_new_equation}
\end{equation}
\begin{equation}
    \frac{d(GPE_r)}{dt} = W_{r,mixing} = W_{r,laminar} + W_{r,turbulent} ,
    \label{gper_new_equation}
\end{equation}
\begin{equation}
    \frac{d(IE_0)}{dt} \approx D(KE) + D(APE) = D_{total} ,
    \label{ieo_equation}
\end{equation}
\begin{equation}
    \frac{d(IE_{exergy})}{dt} \approx - W_{r,mixing} = - W_{r,laminar} -
    W_{r,turbulent} .
    \label{iex_equation}
\end{equation}
In this model, the first three equations are just a rewriting of 
Eqs. (\ref{KE_equation1}) and (\ref{GPE_equation1}), so that the
main novelty is associated with Eq. (\ref{ieo_equation}) and
(\ref{iex_equation}). Physically, Eq. (\ref{ieo_equation}) states
that the viscous and diffusive dissipation processes 
$D(KE)$ and $D(APE)$ mostly affect $T_0$ but not $T_r$, while 
Eq. (\ref{iex_equation}) states that the $IE/GPE_r$ conversion
reduces $IE_{exergy}$ as well as smoothes out $T_r$. The empirical
verification of the validity of the
above equations is the main topic of Section 2.

\subsection{Link with the ocean heat engine controversy}

  In the oceans, turbulent diapycnal mixing is a crucial process, as it
is required to transport heat
from the surface equatorial regions down to the depths cooled by
high-latitude deep water formation. In the traditional picture found in
most oceanography textbooks, 
turbulent diapycnal mixing and deep water formation are usually described
as part of the buoyancy-driven component of the large-scale ocean
circulation responsible for the oceanic poleward transport of heat, often
called the meridional overturning circulation (MOC). Physically, the MOC
is often equated with the longitudinally-averaged circulation taking 
place in the latitudinal/vertical plane. The possible dependence of
the buoyancy-driven circulation on mechanical forcing, which one might
expect in a system as nonlinear as the oceans, has been usually ignored.
However, the idea of a buoyancy-driven circulation unaffected by mechanical
forcing physically makes sense only if one can establish that the mechanical
stirring required to sustain turbulent diapycnal mixing is driven by 
surface buoyancy fluxes. \cite{Munk1998} (MW98 thereafter) questioned 
this view, and argued instead that turbulent diapycnal mixing must in
fact be primarily driven by the wind and tides, and hence that the
buoyancy-driven circulation must in fact be mechanically-controlled. 
Moreover, MW98 analysed the $GPE$ budget of the oceans to derive the
following constraint:
\begin{equation}
         G(KE) = \frac{W_{r,forcing}}{\gamma_{mixing}} ,
        \label{MW98_constraint}
\end{equation}
linking the work rate $G(KE)$ done by the mechanical sources of stirring,
the rate $W_{r,forcing}$ at which high-latitude cooling depletes
$GPE_r$, and the oceanic bulk ``mixing efficiency'' $\gamma_{mixing}$
(or more accurately, the dissipations ratio, as argued previously).
Physically, Eq. (\ref{MW98_constraint}) states that  
the fraction $\gamma_{mixing}$ of $G(KE)$ has to be expanded into turbulent
mixing to raise $GPE_r$ at the same rate $W_{r,forcing}$ at which it is lost.
By using the values $W_{r,forcing}\approx 
0.4\,{\rm TW}$ and $\gamma_{mixing}=0.2$, MW98 concluded that 
$G(KE) = O(2\,{\rm TW})$ is approximately required to sustain the observed
oceanic rates of turbulent diapycnal mixing. This result caused much
stirring in the ocean community, because the wind supplies only about 
$1\,{\rm TW}$, leaving an apparent shortfall of $1\,{\rm TW}$ to close 
the energy budget. This led MW98 to argue that the only plausible candidate
to account for the missing stirring should be the tides, spawning 
a considerable research effort over the past 10 years on the issue of 
tidal mixing.

\par

  Although MW98's arguments have been echoed favourably within the ocean
community, \eg 
\cite{Huang2004,Paparella2002,Wunsch2004,Kuhlbrodt2007,Nycander2007}, 
it remains unclear
why the surface buoyancy fluxes should not be important 
in stirring and driving the oceans, given that the work rate done by
the surface buoyancy fluxes, as measured by the $APE$ production rate,
was previously estimated by \cite{Oort1994} to be 
$G(APE)=1.2\,\pm 0.7\,{\rm TW}$ and hence comparable in importance to
the work rate done by the mechanical forcing. In their paper, MW98 
rather summarily dismissed \cite{Oort1994}'s results by contending that
the so-called 
\cite{Sandstrom1908}'s ``theorem''
requires that $G(APE)$ be negligible, and hence that the buoyancy forcing
cannot produce any significant work in the oceans, but given the highly
controversial nature of \cite{Sandstrom1908}'s paper in the physical 
oceanography community, and its apparent refutation by \cite{Jeffreys1925},
it would seem important to have a more solid physical basis to make any
definitive statements about $G(APE)$. Note that Sandstrom's paper was recently
translated by \cite{Kuhlbrodt2008}, who argues that Sandstrom's did not
initially formulate his results as a theorem, but rather as an inference.
Ascertaining whether $G(APE)$ is
large or small is obviously
crucial to determine whether the MOC is effectively driven by the 
turbulent mixing powered by the winds and tides, as argued by MW98,
or whether it is in fact predominantly buoyancy-driven, as it appears
to be possible if $G(APE)$ is as large as predicted by \cite{Oort1994}.
Further clarification is also needed to understand the possible importance
of some effects neglected by MW98, such as those due to a nonlinear equation
of state, which \cite{Gnanadesikan2005} argue lead to a significant 
underestimation of $G(KE)$, or due to entrainment effects, which 
\cite{Hughes2006} argue lead to a possible significant overestimation of
$G(KE)$.

\subsection{Purpose and organisation of the paper}

  The primary objective of this paper is to clarify the nature of
the energy conversions taking place in turbulent stratified fluids,
with the aim of clarifying 
the underlying assumptions entering MW98's energy constraint 
Eq. (\ref{MW98_constraint}). 
The backbone of the paper are the theoretical derivations presented
in the appendices A and B, which provide a rigorous theoretical support
to understand the links between stirring and irreversible mixing in 
mechanically and thermodynamically-forced thermally-stratified fluids.
Appendix A offers a new derivation of 
\cite{Winters1995}'s framework, which is further extended to the case
of a Boussinesq fluid with a thermal expansion increasing with temperature.
Appendix B is a further extension to the
case of a fully compressible thermally-stratified fluid, in which 
the decomposition of internal energy into three distinct sub-reservoirs
is presented. Section 2 seeks to illustrate the differences between $D(APE)$
and $W_{r,turbulent}$ using a number of different viewpoints, and examine
some of its consequences, in the context of freely decaying turbulence.
Section 3 revisits the issues pertaining to MW98's energy constraint.
Section 4 offers a summary and discussion of the results.

\begin{figure}
\centering
\includegraphics[width=9cm]{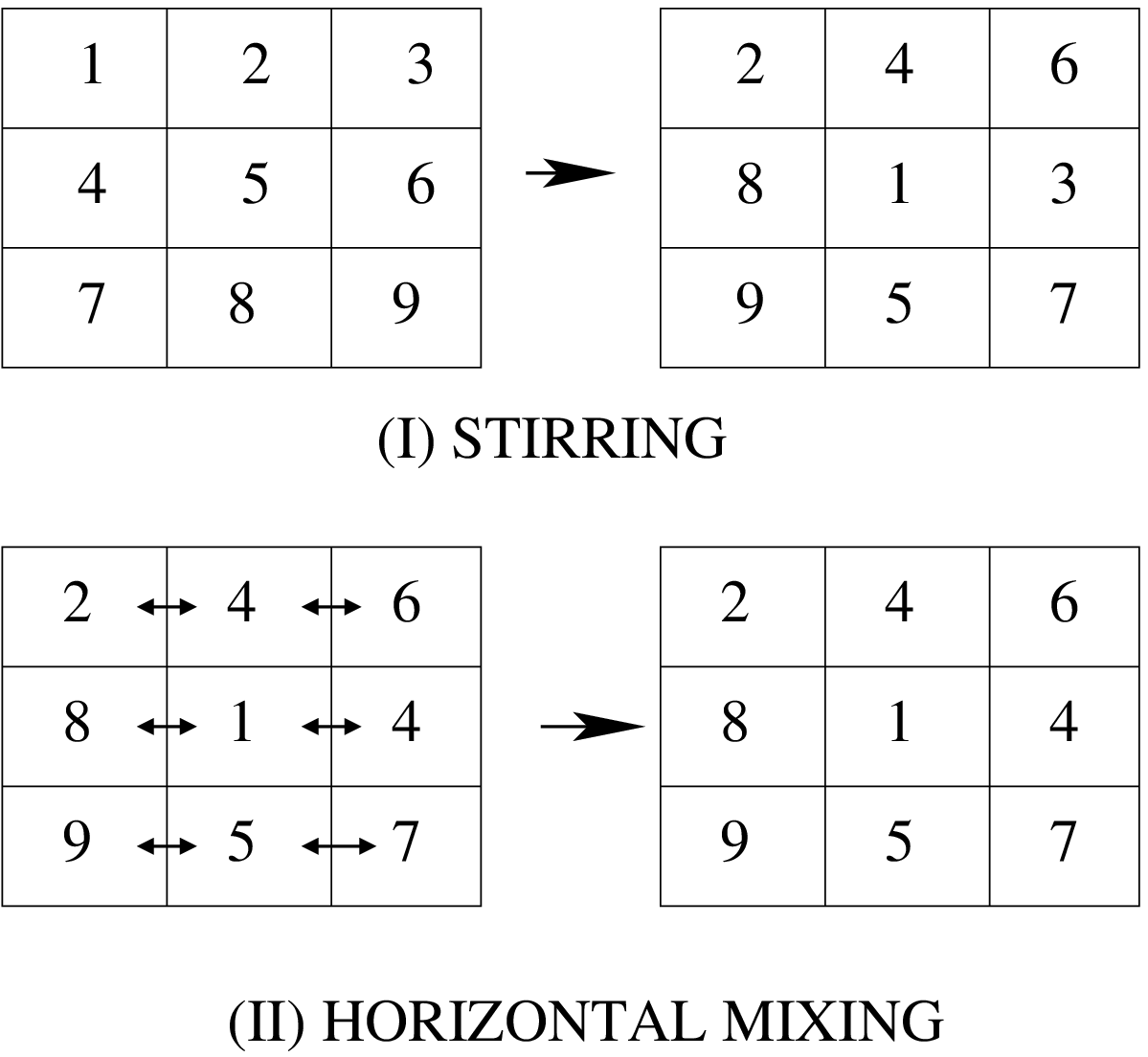}
\caption{Idealised depiction of the numerical experimental protocol used to 
construct Fig. \ref{shuffling_figure}, as well as underlying the 
method for constructing Figs. \ref{bb_figure} and \ref{wr_figure}.
In the top panel, a piece of stratified fluid is cut into pieces
of equal mass that are numbered from $1$ to $N_{tot}$, where 
$N_{tot}$ is the total number of parcels. A random permutation is
generated as a way to shuffle the parcels randomly and adiabatically,
in order to mimic the stirring process. In the bottom panel, all the
parcels lying at the same level are homogenised to the same temperature
by conserving the total energy of the system, which mimics the 
horizontal mixing step illustrated in Fig. \ref{diffusiveroute}.}
\label{stirringmixing}
\end{figure}

\begin{figure}
\centering
\includegraphics[width=13cm]{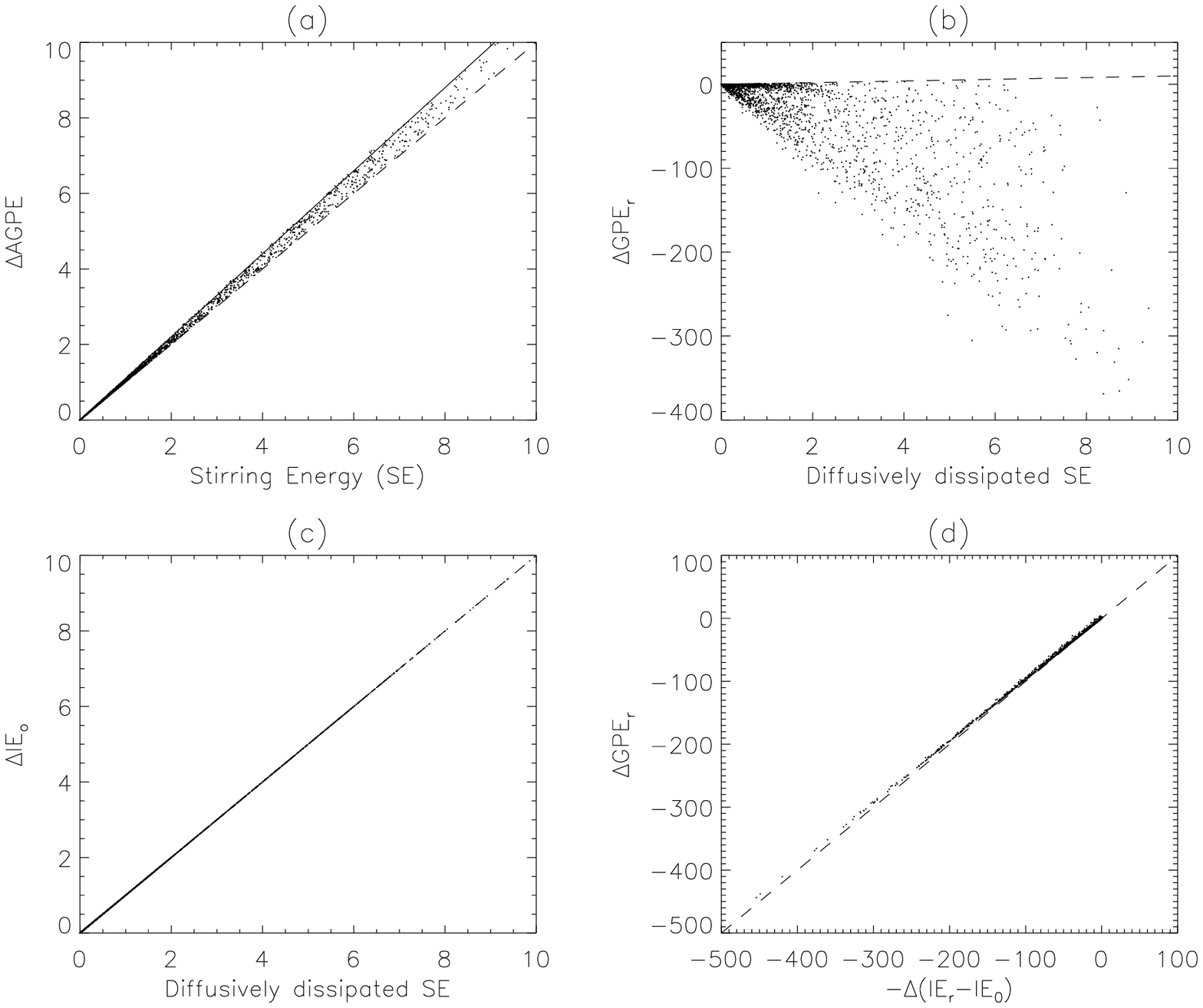}
\caption{(a) The increase of AGPE as a function of the stirring energy SE
(see text for details). Each point represents a different stratification
shuffled by a different random permutation. The continuous line represents
the straight-line of equation $\Delta AGPE = SE$ which would describe the
energetics of the stirring process if $AIE$ could be neglected; 
(b) the change of $GPE_r$ as a function of
the stirring energy SE dissipated by diffusive mixing; the dotted line
is the straight line of equation $\Delta GPE_r = 
{\rm Diffusive\, dissipated\, SE}$
which would describe the energetics of turbulent mixing if the irreversible
conversion $AGPE \longrightarrow GPE_r$ existed; (c) The change in dead
internal energy $IE_0$ as a function of the diffusively dissipated stirring
energy SE. The dashed line is the straight-line of equation 
$\Delta IE_0 = {\rm diffusively\, dissipated\, SE}$; 
the figure shows a near perfect
correlation; (d) The change in $GPE_r$ as a function of the exergy change.
The dashed line is the straight line of equation $\Delta GPE_r = 
- \Delta IE_{exergy}$. The figure shows again a near perfect correlation.}
\label{shuffling_figure}
\end{figure}

\section{A new view of turbulent mixing energetics in freely 
decaying stratified turbulence}
\label{shuffling_experiment}

\subsection{Boussinesq versus Non-Boussinesq energetics}

  As mentioned above, a central point of this paper is to argue that 
irreversible energy conversions in turbulent stratified fluids are best
understood if internal energy is not regarded as a single energy reservoir,
but as the sum of at least three distinct sub-reservoirs. Obviously,
these nuances are lost in the traditional Boussinesq 
description of turbulent fluids, since the latter lacks an explicit 
representation of internal energy, let alone of its three sub-reservoirs.
This does not mean that the Boussinesq approximation is necessarily
inaccurate or incomplete, but rather that the definitive interpretation
of its energy conversions requires to be checked against the understanding
gained from the study of the fully compressible Navier-Stokes equations.
Such a study was carried out, with the results reported in Appendix B.
In our approach, successive refinements of the energy conversions were 
sought, starting from the $KE/APE/PE_r$ system for which the number of
energy conversions is limited and unambiguous. The second step
was to split $PE_r$ into its $GPE_r$ and $IE_r$ components; the third step
was to 
further split $IE_r$ into its exergy $IE_r-IE_0$ and dead $IE_0$ components.
Finally, the last step was to split $APE$ into its $AIE$ and $AGPE$ 
components. These successive refinements are illustrated in Fig.
\ref{basic_energetics} in Appendix B.
An important outcome of the analysis is that the structure and form of the
$KE/APE/GPE_r$ equations (Eqs. 
(\ref{KE_equation1}-\ref{GPE_equation1})) obtained for the L-Boussinesq model
turn out to be more generally valid
for a fully compressible thermally-stratified fluid, so that one still has:
\begin{equation}
    \frac{d(KE)}{dt} = -C(KE,APE) - D(KE) ,
    \label{KE_again}
\end{equation}
\begin{equation}
    \frac{d(APE)}{dt} = C(KE,APE) - D(APE) ,
     \label{APE_again}
\end{equation}
\begin{equation}
    \frac{d(GPE_r)}{dt} = W_{r,mixing} = W_{r,turbulent} + W_{r,laminar}.
     \label{GPEr_again}
\end{equation}
It can be shown, however, that the explicit expressions for
$C(KE,APE)$, $D(KE)$, and $D(APE)$ differ between the two sets
of equations, see Appendices A and B for the details of these
differences. Based on the numerical simulations detailed in the
following, the most important point is probably that $D(APE)$
appears to be relatively unaffected by the details of the 
equation of state, in contrast to $W_{r,mixing}$, which suggests
that the L-Boussinesq model is able to accurately represent the
irreversible diffusive mixing associated with $D(APE)$.
Moreover, since the internal energy contribution to $APE$ is usually
small for a nearly incompressible fluid, it also follows that the
L-Boussinesq model should also be able to capture the time-averaged
properties of $C(KE,APE)$, since the latter is the difference of
two terms expected to be accurately represented by the L-Boussinesq
model based on the APE equation. The L-Boussinesq model, however, will
in general fail to correctly capture the behaviour of $GPE_r$, unless
the approximation of a linear equation of state
 is accurate enough, as seems to be
the case for compositionally-stratified flows for instance, \eg 
\cite{Dalziel2008}. The above properties help to rationalise why
the L-Boussinesq model appears to perform as well as it often does.

\par

  Being re-assured that there are no fundamental structural differences
between the energetics of the $KE/APE/GPE_r$ system in the Boussinesq 
and compressible NSE, the next step is to clarify the link with internal
energy. One of the main results of this paper, derived in Appendix B,
are the following evolution equations for the dead and exergy components
of internal energy:
\begin{equation}
    \frac{d(IE_0)}{dt} \approx D(KE) + D(APE) ,
    \label{ieo_final}
\end{equation}
\begin{equation}
    \frac{d(IE_{exergy})}{dt}
    \approx - W_{r,mixing}  ,
\label{exergy_final}
\end{equation}
which were obtained by neglecting terms scaling as
$O(\alpha P/(\rho C_p))$, for 
some values of $\alpha$, $P$, $\rho$, and $C_p$ typical of the domain
considered, 
where $\alpha$ is the thermal expansion coefficient, $P$ is the pressure,
$\rho$ is the density, and $C_p$ is the heat capacity at constant pressure.
The important point is that such a parameter is very small for
nearly incompressible fluids. For seawater, for instance, typical
values encountered in laboratory experiments done at atmospheric pressure
are $\alpha = 2.10^{-4}\,{\rm K}^{-1}$, $P=10^{5}\,{\rm P_a}$, 
$C_p = 4.10^{3}\,{\rm J.K^{-1}.kg^{-1}}$, $\rho=10^3\,{\rm m^3.kg^{-1}}$,
which yield $\alpha P/(\rho C_p)=5.10^{-6}$. In the deep oceans, this 
value can increase up to $O(10^{-3})$, but this is still very small.
Eqs. (\ref{ieo_final}) and (\ref{exergy_final}) confirm that $D(APE)$
and $D(KE)$ are fundamentally similar dissipative processes, in that they
both convert $APE$ and $KE$ into dead internal energy, while also confirming
that $W_{r,mixing}$ represent a conversion between $IE_{exergy}$ 
and $GPE_r$ both
in the laminar and turbulent cases.

\subsection{Analysis of idealised turbulent mixing events}

  To gain insight into the differences between $D(APE)$
and $W_{r,turbulent}$, the energy budget of an hypothetical turbulent 
mixing event associated with shear flow instability
is examined in the light of Eqs. (\ref{KE_again}-\ref{exergy_final}).
Typically, such events can be assumed to
evolve from laminar conditions with no $APE$.
Once the instability is triggered, $APE$ starts to increase and oscillate 
until the instability subsides and the fluid re-laminarises, 
at which point $APE$
returns to zero. The mixing event causes the shear flow to lose
a certain amount of kinetic energy $|\Delta KE|$, as well as $GPE_r$ to
increase by a certain amount $\Delta GPE_r$, as the result of the 
partial smoothing out of the mean vertical temperature gradient by
molecular diffusion. Integrating
Eqs. (\ref{KE_again}-\ref{exergy_final}) over the duration
of the mixing event yields:
\begin{equation}
    \Delta KE = - \overline{C(KE,APE)} - \overline{D(KE)} ,
\end{equation}
\begin{equation}
   \Delta APE = 0 = \overline{C(KE,APE)} - \overline{D(APE)},
\end{equation}
\begin{equation}
   \Delta GPE_r = \overline{W}_{r,turbulent} + 
    \overline{W}_{r,laminar} ,
\end{equation}
\begin{equation}
   \Delta IE_0 = \overline{D(APE)} + \overline{D(KE)} ,
\end{equation}
\begin{equation}
   \Delta IE_{exergy} = -[\overline{W}_{r,laminar}  
    + \overline{W}_{r,turbulent}],
\end{equation}
where $\Delta (.)$ and the overbar denote a quantity's net change
over the time interval and its time-integrated value respectively. 
From an observational viewpoint, energy conversion terms such as
$\overline{C(KE,APE)}$ are difficult to measure directly; moreover,
the results can be ambiguous, \eg \cite{Barry2001} and references
therein. As a result, energy conversions are probably best inferred
from measuring changes in the different energy reservoirs, as this 
appears to be easier to do accurately. Multiple possible inferences arise,
however, if $IE$ variations are not separated into their $IE_{exergy}$
and $IE_0$ components. Fig. \ref{new_interpretation} illustrates this
point for an hypothetical turbulent mixing experiment with hypothetical
plausible numbers, 
by showing that a given observed net change in $IE$ of 
$+0.79$ units can potentially be explained --- in absence of any
knowledge about the respective variations in $IE_0$ and $IE_{exergy}$
--- as either due to the conversion of $0.8$ unit of $KE$ into $IE_0$ 
minus the conversion of $0.01$ unit of $IE_{exergy}$ into $GPE_r$, or
by the conversion of $0.8$ unit of $KE$ into $IE_0$ plus the conversion
of $0.2$ unit of $APE$ into $IE_0$
minus the conversion of $0.21$ unit of $IE_{exergy}$ into $GPE_r$.
Although the first interpretation is the one implicit in \cite{Winters1995}
and currently favoured in the literature,
it is not possible, based on energy conservation alone, to reject the 
second interpretation. In fact, the only way to discriminate between
the two interpretations requires separately measuring $IE_0$ and
$IE_{exergy}$ variations, as only then are the two interpretations
mutually exclusive.

\subsection{An idealised numerical experimental protocol 
to test the two interpretations}

  To compute $IE_0$ and $IE_{exergy}$, a knowledge of the temperature field
and of the Gibbs function for the fluid considered (see \cite{Feistel2003}
in the case of water or seawater) is in principle sufficient.
It is hoped, therefore,
that the present study can stimulate laboratory measurements of $IE_0$ and
$IE_{exergy}$, in order to provide experimental support (or refutation, as
the case may be), for the present ideas. In the meantime, numerical methods
are probably the only way to assess the accuracy of 
the two key formula Eqs. (\ref{ieo_final}) and (\ref{exergy_final}),
which physically argue that: 1)
the diffusively dissipated $APE$ is nearly entirely
converted into $IE_0$;
and 2) that $GPE_r$ variations are nearly entirely accounted for by 
corresponding variations in $IE_{exergy}$. 
\par

To prove our point, energetically consistent idealised
mixing events are constructed and studied numerically.  
The procedure is as follows. One starts from a piece of thermally stratified
fluid initially lying in its \cite{Lorenz1955}'s reference state in a 
two-dimensional container with a flat bottom, vertical walls, and a free
surface exposed at constant atmospheric pressure at its top. The fluid
is then discretised on a rectangular array of dimension $N_x \times N_z$
into discrete fluid elements
having all the same mass ${\rm \Delta m = \rho \Delta x \Delta z}$,
where $x$ and $z$ are the horizontal and vertical coordinates respectively,
as illustrated in Fig. \ref{stirringmixing}. The initial stratification
has a vertically-dependent temperature profile $T(x,P)=T_r(P)$ regarded
as a function of horizontal position $x$ and pressure $P$. Thousands
of idealised mixing
simulations are then generated according to the following procedure:

\begin{enumerate}

\item {\em Initialisation of the reference stratification.} The initial
stratification is discretised as $T_{i,k} = T(x_i,P_k) = T_r(P_k)$, with
$x_i = (i-1) \Delta x, \,\,i=1,\dots N_x$ and $P_k = P_{min}
+ (k-1) g \Delta m, \,\,k=1,\dots N_z$, where $P_{min}$ 
and $T_r(P_k),k=1,\dots N_z$ are
random generated numbers such that $T_{min}\le T_r(P_k)\le
T_{max}$ that have been re-ordered in the vertical to create a statically
stable stratification, for randomly generated $T_{min}$, $T_{max}$, and
$P_{min}$.
 
\item {\em Random stirring of the fluid parcels.} The fluid parcels are then
numbered from $1$ to $N=N_x \times N_z$, and randomly shuffled by generating
a random perturbation of $N$ elements, such that each parcel conserves its
entropy in the re-arrangement. Such a step is intended to mimic the 
adiabatic stirring of the parcels associated with the $KE \longrightarrow APE$
conversion. 
The random stirring of the fluid parcels requires an external amount of
energy --- called the stirring energy $SE$ --- which is diagnosed by
computing the difference in potential energy between the shuffled state
and initial state, i.e.,
\begin{equation}
       SE = (GPE + IE)_{shuffled} - (GPE+IE)_{initial} .
\end{equation}
The latter computation requires a knowledge of the thermodynamic properties
of the fluid parcels. In this paper, such properties were estimated from
the Gibbs function for seawater of \cite{Feistel2003} by specifying a 
constant value of salinity. Thermodynamic properties such as internal energy,
enthalpy, density, entropy, chemical potential, speed of sound, thermal
expansion, haline contraction, and several others, are easily estimated
by computing partial derivatives with respect to
temperature, pressure, salinity, or any combination thereof, of the
Gibbs function.
The stirring energy $SE$ is none other than \cite{Lorenz1955}'s $APE$
of the shuffled state. Since the stirring leaves the background potential
energy unaffected, the energetics of the random shuffling is given by:
\begin{equation}
       \Delta APE = \Delta AGPE + \Delta AIE = SE ,
       \label{ape_se}
\end{equation}
\begin{equation}
        \Delta GPE_r = \Delta IE_r = 0 ,
        \label{dgper}
\end{equation}
where Eq. (\ref{ape_se}) states that the stirring energy $SE$ is entirely
converted into $APE$, while Eq. (\ref{dgper}) expresses the result that
being a purely adiabatic process, the stirring leaves the background 
reference quantities unaltered. Fig. \ref{shuffling_figure} (a) depicts
$\Delta AGPE$ as a function of $SE$ for thousands of experiments, all
appearing as one particular point on the plot. According to this figure,
$\Delta AGPE$ approximates $\Delta APE$ within about $10\%$.
This illustrates
the point that {\em for adiabatic processes}, $APE$ is well approximated
by its gravitational potential energy component, as expressed by
Eq. (\ref{adiabatic_transfo}).

\item {\em Isobaric irreversible mixing of the fluid parcels.} In the last
step, all the fluid parcels lying in the same layer are mixed uniformly to
the same temperature, by assuming an isobaric process that conserves the
total enthalpy of each layer. Such a process converts a fraction $q SE$ of
the $APE$ into the background $PE_r$, according to:
\begin{equation}
       \Delta APE = \Delta AGPE + \Delta AIE = - q SE
\end{equation}
\begin{equation}
       \Delta GPE_r + \Delta IE_r = q SE
\end{equation}
where $0 < q \le 1$. The factor $q$ is needed here because mixing each
layer uniformly does not necessarily lead to a statically stable
stratification; when this happens, the resulting stratification still 
contains some $APE=(1-q)SE$ associated with the static instability,
so that $q=1$ only when the mixed density profile is statically stable.
\par 
  The change in $\Delta GPE_r$ resulting from the irreversible mixing step
is depicted as a function of the diffusively dissipated stirring energy
$q SE$ in the panel (b) of Fig. (\ref{shuffling_figure}).
If the stirring energy were entirely dissipated
into $GPE_r$, as is classically assumed, then all points should lie on
the line of equation $\Delta GPE_r = q SE$ appearing as the dashed line
in the figure. Even though such a relation appears to work well in a number
of cases, the vast majority of the simulated points corresponds to cases
where $\Delta GPE_r$ is significantly smaller than $q SE$, and even often
negative as expected in the Fofonoff regime discussed above. On the other
hand, if one plots $\Delta IE_0$ as a function of the diffusively dissipated
stirring energy $q SE$, as well as $\Delta GPE_r$ as a function of the
exergy change $\Delta IE_{exergy} = -\Delta (IE_r-IE_0)$, 
as done in panels (c) and
(d) of Fig. (\ref{shuffling_figure}) respectively, then a visually near
perfect correlation in both cases is obtained. This is consistent with
the following relations:
\begin{equation}
         \Delta IE_0 \approx q SE ,
          \label{ieoqse}
\end{equation}
\begin{equation}
         \Delta GPE_r \approx - \Delta (IE_r - IE_0) ,
\end{equation}
and hence in agreement with the approximate Eqs. (\ref{ieo_final}) and
(\ref{exergy_final}). Eq. (\ref{ieoqse}) empirically verifies the
Eq. (\ref{diabatic_transfo}). QED.

\end{enumerate}

\begin{figure}
\centering
\includegraphics[width=12cm]{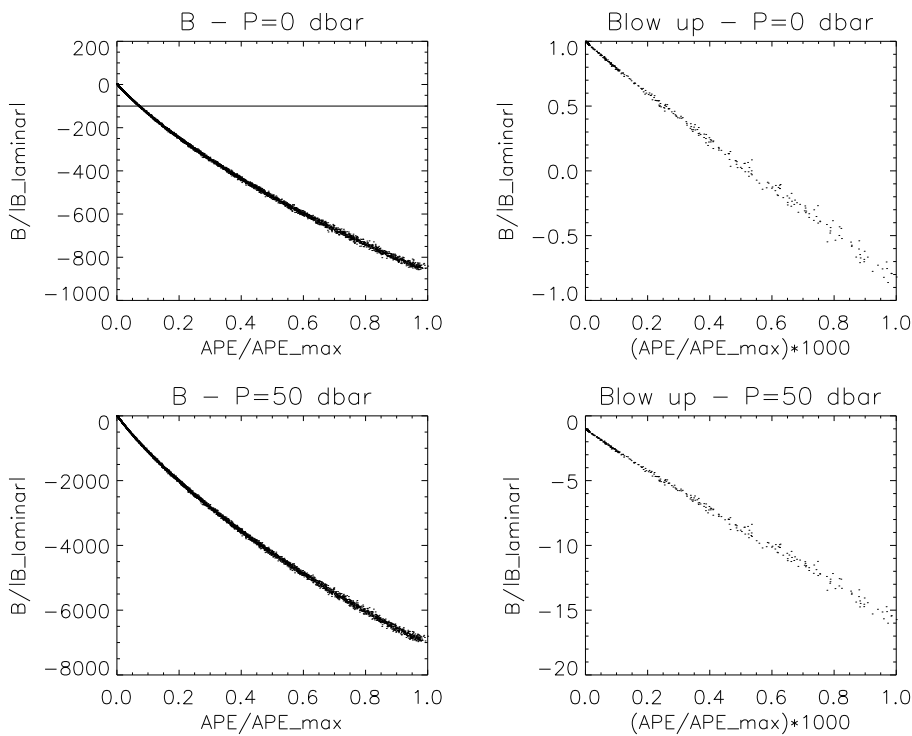}
\caption{(Top panels) The work of expansion/contraction $B$ normalised
by its laminar value (obtained for APE=0)
as a function of a normalised $APE$ for a particular
stratification corresponding to the classical regime, with the right panel
being a blow-up of the left panel. (Bottom panels) Same as above figure,
for the same temperature stratification, but taken at a mean pressure 
of $50\,{\rm dbar}$ instead of atmospheric pressure, which is sufficient
to put the system in the Fofonoff regime. The figures show that although
$B$ is usually negative in every case, it is nevertheless positive for
small values of $APE$ in the classical regime, as expected from
L-Boussinesq theory. The normalisation constant $APE_{max}$
corresponds to the overall maximum of $APE$ for all experiments.}
\label{bb_figure}
\end{figure}

\subsection{Numerical estimates of $B$, $W_{r,mixing}$ and $D(APE)$ as 
a function of $APE$}

  Having clarified the nature of the net energy conversions occurring
in idealised mixing events, the following turns to the estimation of the
turbulent rates of the three important conversion terms $B$, $W_{r,mixing}$, 
and $D(APE)$, which are the main three terms affected by molecular 
diffusion in the fully compressible Navier-Stokes equations, where $B$
is the work of expansion/contraction. As pointed
out in the introduction, enhanced rates fundamentally arise from 
turbulent fluids possessing large amounts of small-scale $APE$. For this
reason, this paragraph seeks to understand how the values of $B$, 
$W_{r,mixing}$ and $D(APE)$ are controlled by the magnitude of $APE$.

\par

  We first focus on the work of expansion/contraction $B$, which takes
the following form:
\begin{equation}
     B = \int_{V} \frac{\alpha P}{\rho C_p} \nabla \cdot 
    \left (\kappa \rho C_p \nabla T \right ) \,dV 
  + \int_{V} \frac{\alpha P}{\rho C_p} \rho \varepsilon \,dV
  - \int_{V} \frac{P}{\rho c_s^2} \frac{DP}{Dt} \,dV ,
   \label{B_exact}
\end{equation}
obtained by regarding $\rho$ as a function of temperature and pressure.
The part of $B$ affected by molecular diffusion is the first term in 
the right-hand side of Eq. (\ref{B_exact}), and the one under focus here.
The second and third term in the r.h.s. of Eq. (\ref{B_exact}) are
respectively caused by the work of expansion due to the viscous dissipation
Joule heating, and to the adiabatic work of expansion/contraction.
The study of these two terms is beyond the scope of this paper.

\par

  The diffusive part of $B$ was estimated numerically for thousands of
randomly generated stratifications, similarly as in the previous paragraph,
using a standard finite difference discretisation of the molecular diffusion
operator. Unlike in the previous paragraph, however, all the randomly 
generated stratifications were computed from only two different reference
states pertaining to the classical and Fofonoff regimes respectively, the
results being depicted in the top and bottom panels of 
Fig. \ref{bb_figure} (with the right panels providing a blow-up of the left 
panels). The main result here is that finite values of $APE$ can make the
diffusive part of $B$ negative and 
considerably larger by several orders of magnitude
than in the laminar $APE=0$ case. This result is important, because it is in
stark contrast to what is usually assumed for nearly incompressible fluids
at low Mach numbers. From Fig. \ref{bb_figure}, it is tempting to conclude
that there exists a well-defined relationship between the diffusive part of
$B$ and $APE$, but in fact, the curve $B=B(APE)$ is more likely to represent
the maximum value achievable by $B$ for a given value of $APE$. Indeed, it
is important to realize that a given value of $APE$ can correspond to widely
different spectral distributions of the temperature field. In the present
case, it turns out that the random generator used tended to generate 
temperature fields with maximum power at small scales, which in turns tend
to maximise the value of $B$ for a particular value of $APE$. 
For the same value of $APE$, smoother stratifications exist with
values of $B$ lying in between the $x$-axis and 
the empirical curve $B=B(APE)$, the latter being expected to depend on
the numerical grid resolution employed.
Nevertheless, Fig. \ref{bb_figure} raises the interesting question of
whether the empirical curve $B=B(APE)$ could in fact describe the behaviour
of the fully developed turbulent regime, an issue that could be explored
using direct numerical simulations of turbulence.

\begin{figure}
\centering
\includegraphics[width=13cm]{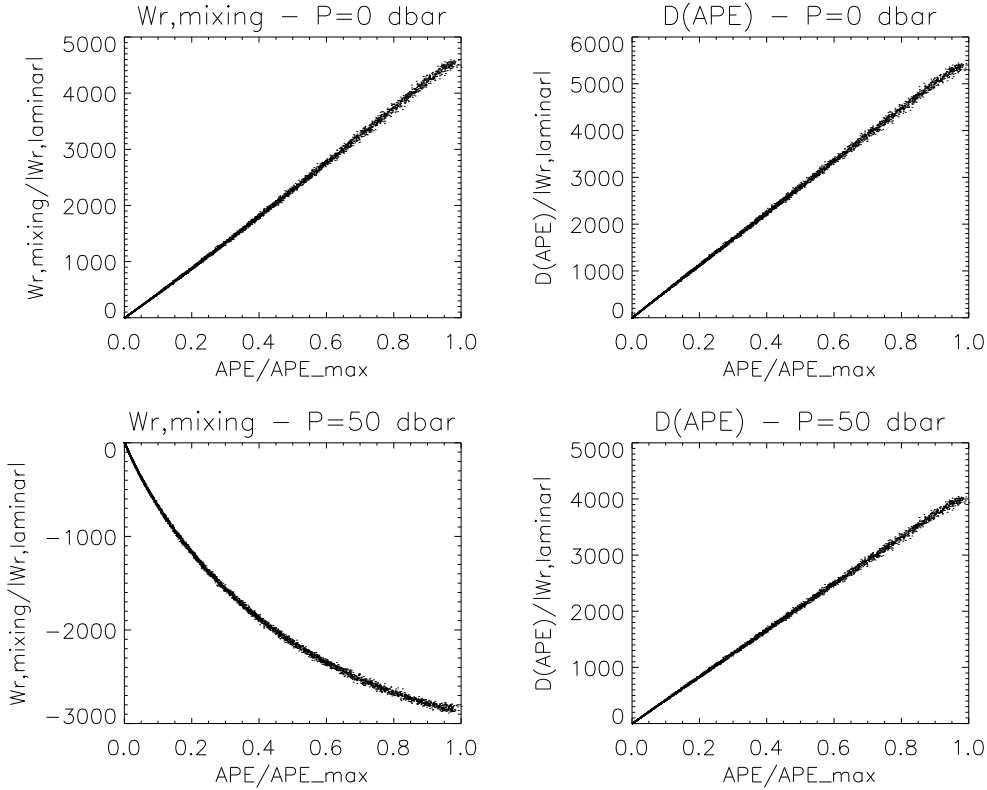}
\caption{(Left panels) The rate of change of $GPE_r$ normalised by its
laminar value as a function of normalised $APE$, in the classical regime
(top panel) as well as in the Fofonoff regime (bottom panel). The 
stratification is identical to that of Fig. \ref{bb_figure}. 
(Right panels) The rate of diffusive dissipation of $APE$ normalised by
$W_{r,mixing}$ laminar value, as a function of a normalised $APE$, in the
classical regime (top panel), as well as for the Fofonoff regime (bottom
panel). The figure illustrates the fact that if the former can be regarded as
a good proxy for the latter in the classical regime, as is usually 
assumed, this is clearly not the case in the Fofonoff regime. The two
figures also illustrate the fact that the former always underestimate the
latter for a thermally stratified fluid, so that observed values of
mixing-efficiencies obtained from measuring $GPE_r$ variations are 
necessarily lower-bounds for actual mixing efficiencies.}
\label{wr_figure}
\end{figure}

  The two remaining quantities of interest are $W_{r,mixing}$ and $D(APE)$,
which were numerically estimated from the following expressions
derived in Appendix B:
\begin{equation}
    W_{r,mixing} = \int_{V} \frac{\alpha_r P_r}{\rho_r C_{pr}} 
     \nabla \cdot \left ( \kappa \rho C_p \nabla T \right )\,dV
   \label{wr_equation}
\end{equation}
\begin{equation}
    D(APE) = \int_{V} \frac{T_r-T}{T} \nabla \cdot \left ( 
      \kappa \rho C_p \nabla T \right ) \,dV
      = \int_{V} \kappa \rho C_p \nabla T \cdot 
        \left ( \frac{T-T_r}{T} \right ) \,dV .
\end{equation}
As for $B$, these two quantities were evaluated for thousands of randomly
generated stratifications as functions of $APE$, starting from the same
reference states as before. The results for $W_{r,mixing}$ are depicted
in the left panels of Fig. (\ref{wr_figure}), and the results for $D(APE)$
in the right panels of the same Figure, with the top and bottom panels
corresponding to the classical and Fofonoff regimes respectively.
The purpose of the comparison is to demonstrate that whereas there exist
stratifications for which the two rates $D(APE)$ and $W_{r,mixing}$ are 
nearly identical (top panel, classical regime), as is expected from the
classical literature about turbulent stratified mixing,
it is also very easy to construct specific cases occurring in the oceans
for which the
two rates become of different signs (bottom panel, Fofonoff regime).
The other important result is the relative insensitivity of $D(APE)$
to the nonlinear character of the equation of state compared to 
$W_{r,mixing}$, suggesting that the use of the L-Boussinesq model can
still accurately describe the $KE/APE$ interactions even for strongly
nonlinear equations of states, although it would fail to do a good job
of simulating the evolution of $GPE_r$ outside the linear equation of
state regime. This also suggests that the L-Boussinesq should be adequate
enough to study the mixing efficiency of turbulent mixing events over a
wide range of circumstances, provided
that by mixing efficiency one means the quantity
$\gamma_{mixing}=D(APE)/D(KE)$, and not $\gamma_{mixing}
=W_{r,turbulent}/D(KE)$.
Finally, we also experimentally verified (not shown) that $D(APE)$
is well approximated by the quantity $W_{r,mixing}-B$, as is expected
when $AIE$ is only a small fraction of $APE$.

\begin{figure}
\centering
\includegraphics[width=13cm]{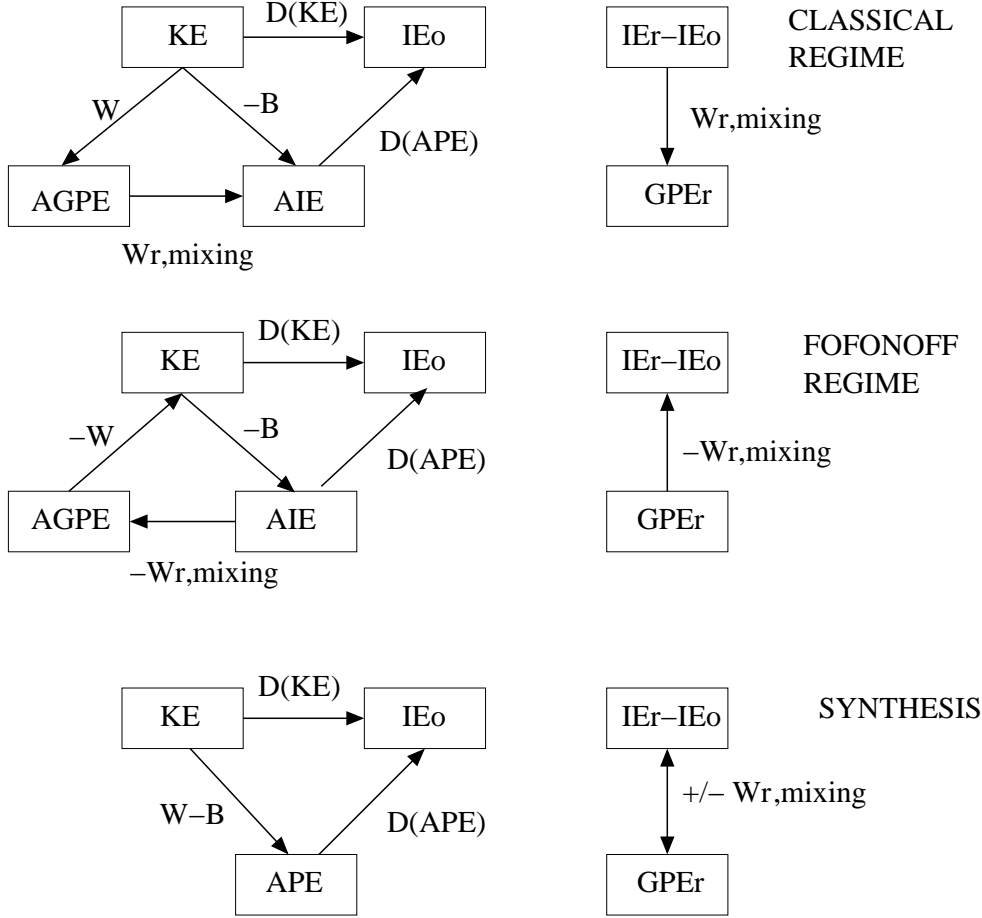}
\caption{The energetics of freely decaying turbulence for the classical
regime (top panel), the Fofonoff regime (middle) panel, and a synthesis
of both regimes obtained by subsuming $AGPE$ and $AIE$ into $APE$ alone.
Note the similarity of the energetics in the lower panel and that of the
re-interpreted Boussinesq energetics of the lower panels of 
Fig. \ref{new_interpretation}.}
\label{decaying1}
\end{figure}

\subsection{Synthesis}
 
  The energetics of freely decaying turbulence is summarised in 
Fig. \ref{decaying1} for the classical (top panel) and 
Fofonoff (middle panel) regimes, with the bottom panel attempting further
synthesis by combining $AIE$ and $AGPE$ into a single reservoir for $APE$,
and the two regimes into a single diagram. Doing so
makes the bottom panel of Fig. (\ref{decaying1}) basically identical to
the Boussinesq energy flowchart depicted in the bottom panel of
Fig. \ref{new_interpretation}. Interestingly,
the middle panel suggests that the Fofonoff regime may differ from the
extensively studied classical regime in several fundamental ways. Indeed,
whereas both $W$ and $B$ act as net sinks of $KE$ in the classical regime,
it appears possible in the Fofonoff regime for some fraction of the $KE$ 
dissipated into $AIE$ to be recycled back to $KE$. This is reminiscent
of the positive feedback on the turbulent kinetic energy discussed by
\cite{Fofonoff1998,Fofonoff2001}, who suggested that such a feedback
would enhance turbulent mixing and hence speed up the return to the
classical regime after sufficient reduction of the vertical temperature
gradient. If real, such a mechanism would be very important to study and
understand, as potentially providing a limiting process on the maximum
value achievable by the buoyancy frequency, with important implications
for numerical ocean models parameterisations. In his papers, however,
Fofonoff envisioned the positive feedback on turbulent $KE$ as being 
associated with the conversion of $GPE_r$ into $AGPE$, but this goes 
against the findings of this paper arguing that $GPE_r$ can only be 
exchanged with the exergy reservoir. Fofonoff's feedback mechanism was
also criticised by \cite{McDougall2003b} on different grounds. While the
present results do not necessarily rule out Fofonoff's feedback mechanism,
they suggest that the latter probably does not work as originally envisioned
by Fofonoff, if it works at all (\cite{McDougall2003b}'s arguments are
not really conclusive either, as they implicitly rely on the existence
of the $APE/GPE_r$ conversion). In any case, the
issue seems to deserve more attention, given that many places in the oceans
appear to fall into Fofonoff's regime. 

\section{Forced/dissipated balances in the oceans}
\label{heatenginetheory}

\begin{figure}
\centering
\includegraphics[width=11cm]{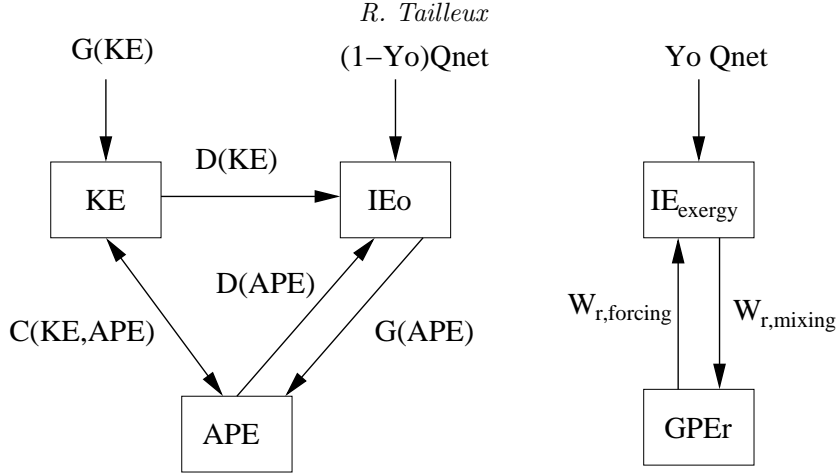}
\caption{Energy flowchart for a mechanically- and buoyancy-driven
thermally-stratified fluid, where $Q_{net}=Q_{heating}-Q_{cooling}$.
At leading order, the ``dynamics'' (the $KE/APE/IE_0$ system) is
decoupled from the ``thermodynamics'' (the $IE_{exergy}/GPE_r$ system).
The dynamics/thermodynamic coupling occurs through the correlation
between $D(APE)$ and $W_{r,mixing}$, as well as through the
correlation between $G(APE)$ and $W_{r,forcing}$.}
\label{forced_energetics}
\end{figure}

\subsection{A new approach to the mechanical energy balance in the oceans}

  Prior to re-visiting MW98's energy constraint Eq. (\ref{MW98_constraint}),
we start by establishing a number of important results for 
mechanically- and thermodynamically-forced thermally-stratified fluids,
based on the results derived in Appendices A and B.
The main modifications brought about
by the mechanical and thermodynamical forcing is the apparition of forcing
terms, i.e., terms involving the external forcing, in the evolution 
equations for $KE$, $APE$, and $GPE_r$ as follows:
\begin{equation}
     \frac{d(KE)}{dt} = -C(KE,APE) - D(KE) + G(KE),
     \label{KE_balance}
\end{equation}
\begin{equation}
     \frac{d(APE)}{dt} = C(KE,APE) - D(APE) + G(APE),
     \label{APE_balance}
\end{equation}
\begin{equation}
     \frac{d(GPE_r)}{dt} = 
    \underbrace{W_{r,turbulent} + W_{r,laminar}}_{W_{r,mixing}}
     - W_{r,forcing} , . 
     \label{PEr_balance}
\end{equation}
where $G(KE)$ is the work rate done by the external mechanical
forcing, $G(APE)$ is the work rate done by the buoyancy forcing,
and $W_{r,forcing}$ is rate of change of $GPE_r$ (usually a loss,
hence the assumed sign convention) due to the buoyancy
forcing. The resulting energy transfers are illustrated in the
energy flowchart depicted in Fig. \ref{forced_energetics}. This
figure shows that at leading order, the ``Dynamics''
--- associated with the reservoirs $KE/APE/IE_0$ --- is 
decoupled from
the ``Thermodynamics'' --- associated with the 
$GPE_r/IE_{exergy}$ energy reservoirs. Indirect coupling occurs,
however, from the fact that $D(APE)$ and $W_{r,turbulent}$ 
on the one hand, and $G(APE)$ and $W_{r,forcing}$ on the other
hand, are strongly correlated to each other.

\subsubsection{Link between $G(APE)$ and $W_{r,forcing}$}

 Unlike in the L-Boussinesq model, $G(APE)$ and $W_{r,forcing}$
differ from each other in a real compressible fluid, for the same
reasons that $D(APE)$ differs from $W_{r,turbulent}$, as is 
apparent from their exact formula given by Eqs. (\ref{gape_definition})
and (\ref{wrforcing_definition}) in Appendix B: 
\begin{equation}
       G(APE) = \int_S \frac{T-T_r}{T} \kappa \rho C_p 
      \nabla T \cdot {\bf n} dS ,
      \label{GAPE_formula}
\end{equation}
\begin{equation}
      W_{r,forcing} = - \int_{S} \frac{\alpha_r (P_r-P_a)}{\rho_r C_{pr}}
      \kappa \rho C_p \nabla T \cdot {\bf n} dS .
      \label{wr_formula}
\end{equation}
In order to understand by how much $W_{r,forcing}$ differs from
$G(APE)$ in a real fluid, it is useful to expand $T$ as a 
Taylor series around $P=P_r$, i.e.,
$$
    T = T(P_a) =  T_r + \Gamma_r (P_a - P_r) + \cdots 
$$
where $\Gamma_r = \alpha_r T_r/(\rho_r C_{pr})$ is the
adiabatic lapse rate, \eg \cite{Feistel2003}. As a result
\begin{equation}
   \frac{T-T_r}{T} = -\frac{\Gamma_r (P_r-P_a)}{T_r} + \cdots
   \approx -\frac{\alpha_r (P_r-P_a)}{\rho_r C_{pr}} .
   \label{ttr}
\end{equation}
Inserting Eq.(\ref{ttr}) into Eq. (\ref{GAPE_formula}) reveals
that $G(APE)$ and $W_{r,forcing}$ are in fact equal at leading
order. In that case, therefore, the equality between
$G(APE)$ and $W_{r,forcing}$ that exactly holds in the L-Boussinesq
model appears to be a much better approximation than the 
corresponding equality between $D(APE)$ and $W_{r,turbulent}$.
For this reason, we shall neglect the differences between
$G(APE)$ and $W_{r,forcing}$ in the following.

\subsubsection{Steady-state Mechanical energy balance}

\begin{figure}
\centering
\includegraphics[width=13cm]{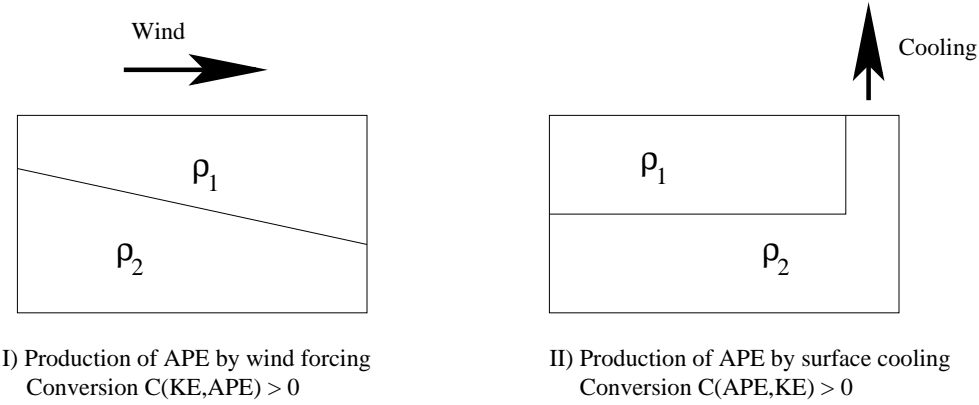}
\caption{Idealised depictions of mechanically-driven (left panel)
and buoyancy-driven (right panel) creation of $APE$. I) A wind blowing 
at the surface of a two-layer fluid causes the tilt of the layer interface,
resulting in a net $C(KE,APE)>0$ conversion. II) Localised cooling at 
high-latitudes sets the density of a fraction of the upper layer to that
of the bottom layer, also inducing a tilt in the layer interface.
The return of the interface to equilibrium conditions (flat interface)
results in a net $C(APE,KE)>0$ conversion.}

\label{basic_conversions}
\end{figure}

 Under steady-state conditions, summing Eqs. 
(\ref{KE_balance}) and (\ref{APE_balance}) yields: 
\begin{equation}
      G(KE) + G(APE) = D(APE) + D(KE) ,
       \label{expected_balance}
\end{equation}
which simply states that in a steady-state, the production of mechanical
energy by the wind and buoyancy forcing is balanced by the viscous and
diffusive dissipations of $KE$ and $APE$ respectively.
Fig. \ref{basic_conversions} schematically illustrates how the wind and
buoyancy forcing can both contribute to the creation of $APE$.
The main novelty here is to make it clear that $D(APE)$ is a ``true''
dissipation mechanism, i.e., one that degrades mechanical energy into
internal energy (as does viscous dissipation), not one converting 
mechanical energy into another form of mechanical energy (i.e., $GPE_r$).
This suggests regarding $D(APE)+D(KE)$ as the total dissipation of
available mechanical energy $ME=KE+APE$.

\par

 In contrast, most studies of ocean energetics of the past decade have 
tended to subsume the $APE$ production and dissipation terms into the 
 single term $B=G(APE)-D(APE)$, in which case Eq. (\ref{expected_balance})
becomes:
\begin{equation}
     G(KE) + B = D(KE) .
    \label{classical_balance}
\end{equation}
The problem in writing the mechanical energy balance under this form
is that it erroneously suggests that
$B$, rather than $G(APE)$, is the work rate done by surface buoyancy fluxes,
and that viscous dissipation is the only form of mechanical energy dissipation.
For instance, \cite{Wang2005} estimated $B=O(1.5\,{\rm GW})$ in the oceans,
in the context of the L-Boussinesq model, which is about three 
orders of magnitude less than the work rate done by the wind
and tides.
In one of the most recent review
about ocean energetics by \cite{Kuhlbrodt2007}, it
is \cite{Wang2005}'s estimate for $B$ that is presented as the work rate
done by surface buoyancy fluxes, while \cite{Oort1994}'s previous result
for $G(APE)$ is regrettably omitted.

\par

 In the L-Boussinesq model, $B$ takes the  
particular form $B=\kappa g (\langle T \rangle_{top}
- \langle T \rangle_{bottom} )$, where $\langle T \rangle_{top}$
and $\langle T \rangle_{bottom}$ are the area-integrated 
surface and bottom temperature respectively,
so that in the absence of mechanical forcing,
Eq. (\ref{classical_balance}) becomes:
\begin{equation}
        \kappa g \left ( \langle T \rangle_{top} -
         \langle T \rangle_{bottom} \right ) = D(KE) .
         \label{ppy02}
\end{equation}
In a study addressing the issue of horizontal convection, recently
reviewed by \cite{Hughes2008},
\cite{Paparella2002} proved that the left-hand side of
(\ref{ppy02}) must be bounded by $\kappa$ times some finite constant 
when the fluid is forced by a surface temperature condition, with 
no-normal flux applying everywhere else. This result is now commonly
referred to as the ``anti-turbulence theorem'', for the bound implies 
that $D(KE)$ must vanish in the ``inviscid'' limit (used here to mean
both vanishing molecular viscosity {\em and} diffusivity), thus violating
the so-called ``zeroth law of turbulence'', an empirical law
grounded in many observations showing that the viscous dissipation of $KE$
in homogeneous turbulent fluid flows remain finite and independent of 
molecular viscosity as the Reynolds number is increased indefinitely.

\subsubsection{Actual implications of the anti-turbulence theorem}

  As shown by \cite{Wang2005}, \cite{Paparella2002}'s bound suggests that
the oceanic viscous dissipation $D(KE)$ would be less than $1.5\,{\rm GW}$
in absence of mechanical forcing. Since this value is several orders of
magnitude than observed oceanic values of $D(KE)$, the result demonstrates
that mechanical forcing is essential to account for the latter. By itself,
however, the result says nothing about whether mechanical forcing is also
essential to account for the observed turbulent rates of diapycnal mixing,
since, as far we are aware, the values of $R_f$ and $\gamma_{mixing}$ for
horizontal convection
have never been determined before. Indeed, the anti-turbulence theorem 
only imposes that the difference $B=G(APE)-D(APE)$ be small, but this does
not forbid $G(APE)$ and $D(APE)$ to be individually very large.
In fact, the current theoretical and numerical evidence suggest that 
$G(APE)$ and $D(APE)$ increase with the Rayleigh number $R_a$. Indeed, this is
suggested by \cite{Paparella2002}'s numerical experiments, which show
the function $\Phi$ (given by Eq. \ref{Phi_function}), which we interpreted
as a measure of the cox number $O(K_T/\kappa)$ (and hence of $D(APE)$)
to increase with 
$R_a = g \alpha \Delta T H^3/(\nu \kappa)$ 
as tabulated in Table \ref{phi_values}. Although such values appear to be
much smaller than observed $O(10^2-10^3)$ Cox numbers, they also correspond
to Rayleigh numbers that are about 13 to 14 orders of magnitude smaller 
than occuring in the oceans, leaving open the possibility for $\Phi$ to
be possibly much larger, possibly as large as encountered in the oceans.
In a related study, \cite{Siggers2004} derived a bound on $\Phi$ (which
they related to a horizontal Nusselt number), which does not exclude the
possibility that horizontal convection, on its own, could support a 
north-south heat transport of the observed magnitude. A further discussion
of the physics of horizontal convection based on laboratory experiments is
provided by \cite{Mullarney2004}. In summary, while
the anti-turbulence theorem demonstrates the need for mechanical forcing
to account for the observed values of kinetic energy dissipation, the 
question of whether mechanical forcing is needed to sustain diapycnal mixing
rates and a north-south heat transport of the observed magnitude is still
largely open. In particular, it is important to point out that although
the anti-turbulence theorem rules out the possibility of elevated values of
kinetic energy dissipation in absence of mechanical forcing, it does not
rule out the possibility of elevated values of diapycnal mixing rates.
In this respect, \cite{Paparella2002}'s suggestion that horizontal
convection should be regarded as ``non-turbulent'' appears somewhat misleading.

\begin{table}
\begin{center}
\begin{tabular}{rlllll}
   $R_a$  & $10^6$ & $2.10^6$ & $3.10^6$ & $4.10^6$ & $5.10^6$ \\
   $\Phi$ & $6.2$ & $6.9$ & $7.5$ & $7.9$ & $8.3$   \\
\end{tabular}
\end{center}
\label{phi_values}
\caption{Values of $\Phi$ as a function of the Rayleigh number $R_a$ 
reproduced from Fig. 4 of \cite{Paparella2002}. Values of $R_a$ 
appropriate to the oceans are of the order $R_a=O(10^{20})$.}
\end{table}

\subsubsection{Back-of-the-envelope 
estimate of $G(APE)$ for the world oceans}

  In order to make progress on the above issue, it is essential to 
determine how large $G(APE)$ can be in the oceans. As mentioned above,
\cite{Oort1994} inferred $G(APE) = 1.2 \pm 0.7\, {\rm TW}$ from observations, 
and hence to be nearly as large as the work rate done by the wind, 
but this estimate was questioned by MW98 on the basis 
of Sandstrom's ``theorem''. A possible source of error in \cite{Oort1994}'s is
its reliance on the so-called Lorenz approximation, which is often said to
overestimate $G(APE)$, \eg \cite{Huang1998}.

\par
 
 In fact, the simplest method to convince oneself that $G(APE)$ must be
large in the oceans comes from the result that $G(APE)\approx W_{r,forcing}$
established previously, which states that if the rate of decrease of $GPE_r$
due to the buoyancy forcing is large, so must it be the case for $G(APE)$.
Since MW98 inferred $W_{r,forcing}\approx 0.4\,{\rm TW}$, one can immediately
conclude $G(APE) \approx 0.4\,{\rm TW}$, which it turns out is close to the
lower bound of \cite{Oort1994}'s estimate, consistent with the idea that the
method used by the latter should overestimate $G(APE)$. This immediately 
establishes that MW98's assumption that $G(APE)$ is small is inconsistent
with their assumption that $W_{r,forcing}$ is large. It also establishes that
Sandstrom's ``theorem'', whatever it means, cannot say anything 
meaningful about $G(APE)$.

\par 

 An independent way to estimate $G(APE)$ is by using the exact
formula for $G(APE)$ or $W_{r,forcing}$ given by Eqs. (\ref{GAPE_formula})
and (\ref{wr_formula}) recalled above.
In these formulae, $T_r$ is the temperature that a surface parcel
would have if lifted adiabatically to its reference level. Since the
oceans are on average heated and cooled where they are the warmest
and coolest respectively, the parcels' reference level will be on
average close to the surface in the warm regions, but much deeper
in the cold regions. Eq. (\ref{GAPE_formula}) must therefore be
dominated by surface cooling. Using the near equality between $G(APE)$
and $W_{r,forcing}$, we take as our estimate for $G(APE)$ the expression:
\begin{equation}
     G(APE) \approx \left ( \frac{\alpha_r P_r}{\rho_r C_{pr}}
     \right )_{cooling} Q_{cooling}
\end{equation}
with a value of $(\alpha_r P_r/(\rho_r C_{pr}))_{cooling}$ appropriate
to the regions of cooling. 
Using the values $\alpha = 5.10^{-5},{\rm K}^{-1}$, $P_r = 2000\,{\rm dbar}
= 2.10^7\,{\rm Pa}$, $\rho_r = 10^3\,{\rm kg}$, $C_{pr}=4.10^3\,
{\rm J.kg^{-1}.K^{-1}}$, and $Q_{cooling}=2\,{\rm PW}$ yields:
$$
     G(APE) = \frac{5.10^{-5}\times 2.10^7}{10^3\times 4.10^3}
      \times 2.10^{15} \,{\rm W} = 0.5\,{\rm TW}
$$
which is very close to \cite{Munk1998}'s estimate for $W_{r,forcing}$,
and is consistent with \cite{Oort1994}'s lower bound for $G(APE)$.
The large value of $G(APE)$ suggests that buoyancy forcing can actively
participate in maintaining turbulent diapycnal mixing in the oceans, 
which should
be reflected by a higher value of $\gamma_{mixing}$ than the value
$\gamma_{mixing}=0.2$ currently used in the literature about the subject.

\subsection{A new look at the $GPE_r$ balance and \cite{Munk1998}'s theory}

  Having clarified the ``available mechanical energy balance'', we now
turn to the $GPE_r$ budget, with the aim of elucidating the assumptions
underlying MW98's constraint on the energy requirement for sustaining
diapycnal mixing in the oceans. In a steady-state, the $GPE_r$ budget
given by Eq. (\ref{PEr_balance}) becomes:
\begin{equation}
      W_{r,mixing} = W_{r,turbulent} + W_{r,laminar} =  W_{r,forcing} ,
      \label{steady_gper}
\end{equation}
where the explicit expressions for $W_{r,mixing}$ and $W_{r,forcing}$
are given at leading order by the following expressions:
\begin{equation}
        W_{r,mixing} \approx \int_{V} \kappa \rho C_p \nabla T \cdot
        \nabla \left ( \frac{\alpha_r T_r}{\rho_r C_{pr}} \right ) dV ,
        \label{Wr_mixing}
\end{equation}
\begin{equation}
        W_{r,forcing} \approx \int_{S} 
      \left ( \frac{\alpha_r P_r}{\rho_r C_{pr}} \right )
        \kappa \rho C_p \nabla T \cdot {\bf n} dS ,
        \label{Wr_forcing}
\end{equation}
which are valid for a fully compressible thermally-stratified ocean.
Note that for the L-Boussinesq model, implicitly considered by MW98,
$W_{r,mixing}$ can be rewritten as:
\begin{equation}
      W_{r,mixing} \approx \int_{V} \kappa 
    \| \nabla z_r \|^2 \alpha_r \frac{\partial T_r}{\partial z_r} dV
     \approx \int_{V} K_T \rho_0 N_r^2 dV
    \label{mw_mixing}
\end{equation}
by using the definition of turbulent diapycnal diffusivity of 
\cite{Winters1995} for $K_T$. 
As Eq. (\ref{mw_mixing}) is exactly the expression
used by MW98, this establishes that MW98's analysis actually pertains to
the $GPE_r$ budget, not the $GPE$ budget, and that their results should
logically follow from Eq. (\ref{steady_gper}).

\par

   In order to show that this is indeed the case, simply use the
definition $\gamma_{mixing}=D(APE)/D(KE)$
in combination with the mechanical energy balance to express $D(APE)$ in terms
of the total mechanical energy input $G(APE)+G(KE)$ as follows:
\begin{equation}
     D(APE) = \frac{\gamma_{mixing}}{1+\gamma_{mixing}} 
    [G(APE)+G(KE)] = R_f [G(APE)+G(KE)],
    \label{dape_forcing}
\end{equation}
where $R_f = \gamma_{mixing}/(1+\gamma_{mixing})$ is the dissipation flux
Richardson number defined in the introduction. 
Now, writing $W_{r,turbulent}=\xi D(APE)$ as proposed in this paper
to account for a nonlinear equation of state, neglecting $W_{r,laminar}$
compared to $W_{r,turbulent}$, and using the result that 
$W_{r,forcing}\approx G(APE)$ demonstrated previously,
Eq. (\ref{steady_gper}) becomes:
\begin{equation}
     \xi D(APE) \approx G(APE) .
     \label{xidape}
\end{equation}
The desired result is obtained by combining Eqs. (\ref{dape_forcing}) and
(\ref{xidape}), which yields:
\begin{equation}
     G(KE) \approx \frac{1+(1-\xi)\gamma_{mixing}}{\xi \gamma_{mixing}} G(APE)
       = \frac{1-\xi R_f}{\xi R_f} G(APE) .
     \label{general_constraint}
\end{equation}
This formula generalises MW98's Eq. (\ref{MW98_constraint}) to account for a
nonlinear equation of state, the effects of which being contained
in the single parameter $\xi<1$. It is easily seen that MW98's Eq.
(\ref{MW98_constraint})
is recovered simply by setting $\xi=1$ in Eq. (\ref{general_constraint}),
using the result that $G(APE)\approx W_{r,forcing}$. This formula is further
extended and discussed in \cite{Tailleux2009b}, in the context of idealized
experiments of mechanically-stirred horizontal convection.

\begin{table}
\begin{center}
   \begin{tabular}{lrrr}
      & $\gamma_{mixing}
   = 0.2$ & $\gamma_{mixing} = 0.5$ & $\gamma_{mixing} = 1$ \\ \\
   $\xi = 1$ & $2\,{\rm TW}$ & $0.8\,{\rm TW}$ & $0.4\,{\rm TW}$ \\
   $\xi = 0.5$ & $4.4\,{\rm TW}$ & $2\,{\rm TW}$ & $1.2\,{\rm TW}$ \\
    \end{tabular}
\end{center}
\label{table_gke}
\caption{Mechanical energy requirements on $G(KE)$
depending on different assumed
values for the dissipations ratio $\gamma_{mixing}$ and nonlinearity parameter
$\xi$, as computed from Eq. (\ref{general_constraint}) using 
$G(APE)=0.4\,{\rm TW}$, in line with \cite{Munk1998}'s 
assumptions.}
\end{table}

  The confirmation that $G(APE)$ should actually
be nearly as large as the work rate done by the mechanical forcing 
makes it possible for buoyancy forcing to drive possibly a very large
fraction of the oceanic turbulent diapycnal mixing, which should be
reflected in an appropriate value for
$\gamma_{mixing}$. As noted in the introduction,
buoyancy-driven turbulent mixing is usually significantly more efficient
than mechanically-driven turbulent mixing, suggesting that a value of
$\gamma_{mixing}$ significantly larger than the value
$\gamma_{mixing}=0.2$ should be used in Eq. (\ref{general_constraint}).
Likewise, the fact that the nonlinear equation of state for seawater
is found to strongly affect changes in $GPE_r$, as showed in the previous
section, suggests that a value of $\xi$ between $0$ and $1$ should be used
(note that $\xi$ cannot be negative if a steady-state is to exist). 
A detailed discussion of which values should actually be used for 
$\gamma_{mixing}$ and $\xi$ is beyond the scope of this paper, however,
as much more needs to be understood about buoyancy-driven and
mechanically-driven turbulent mixing in a non-Boussinesq stratified fluid
before one may become confident enough to speculate on the ``right'' values.
In order to fix ideas, however, it is useful to compute the energy requirement
on turbulent mixing predicted by Eq. (\ref{general_constraint}) for plausible
values of $\xi$ and $\gamma_{mixing}$, as reported in 
Table \ref{table_gke}, under the assumption that
 $W_{r,forcing}\approx G(APE) \approx 0.4\,{\rm TW}$.

\par

 As expected, decreasing $\xi$ at fixed $\gamma_{mixing}$ increases the
requirement on $G(KE)$. This is consistent with \cite{Gnanadesikan2005}'s
conclusions that cabelling (i.e., the contraction upon mixing stemming
from the nonlinear character of the equation of state for seawater) 
increases the requirement on $G(KE)$. Likewise, increasing $\gamma_{mixing}$
at fixed $\xi$ decreases the requirement on $G(KE)$. This is consistent
with \cite{Hughes2006}'s argument that the requirement on $G(KE)$ can
be decreased if the entrainment of ambient water by the sinking cold plumes
is accounted for. Indeed, taking into account entrainment effects is 
equivalent to increasing $\gamma_{mixing}$, since entrainment is physically
associated with buoyancy-driven turbulent mixing, as far as we understand
the issue. Note that decreasing $\xi$ seems to 
require increasing $\gamma_{mixing}$ if one accepts the idea that no
more than $2\,{\rm TW}$ is available from mechanical energy sources to
stir the oceans. Alternatively, one could also perhaps question the assumption
that the oceans are truly in a steady state.

\section{Summary and conclusions}\label{sec:concl}

  In this paper, we extended \cite{Winters1995} APE framework to
the fully compressible Navier-Stokes equations, with the aims of: 
1) clarifying the nature of the energy conversion taking place in turbulent
stratified fluids; 2) clarifying the role of the surface buoyancy fluxes in
\cite{Munk1998}'s constraint on the mechanical sources of stirring required
to sustain diapycnal mixing in the oceans. The most important results 
are that 
the well-known turbulent increase in background $GPE_r$, commonly thought
to occur at the expense of the diffusively dissipated $APE$, actually
occurs at the expense of (the exergy part of) $IE$. On the other hand,
the $APE$ dissipated by molecular diffusion is found to be dissipated
into (the dead part of) $IE$, i.e., the same kind of $IE$ the viscously
dissipated $KE$ is converted into, not into $GPE_r$.
Turbulent stirring, therefore, should not be viewed as introducing a new
form of mechanical-to-mechanical $APE/GPE_r$ conversion, but simply as
enhancing the existing $IE/GPE_r$ conversion rate, 
in addition to enhancing the viscous dissipation rate of $KE$, as well
as the diffusive entropy production and $APE$ dissipation rates.
These results are important, for they significantly alter the current
understanding about the nature of turbulent diapycnal mixing and its links
with the dissipation of mechanical energy and turbulent increase of $GPE_r$.
 In particular, the possibility
that $GPE_r$ may decrease as a result of turbulent mixing, not necessarily
increase as is commonly thought, is to be emphasised. Moreover, the fact
that the turbulent increase of $GPE_r$ is associated with an enhanced 
$IE/GPE_r$ conversion physically implies that compressible effects must be 
considerably larger than previously thought, raising fundamental questions
about the possible limitations of the widely used incompressible assumption
in the modelling of fluid flows at low Mach numbers, which further work
should aim at elucidating. Finally, the 
present results also have implications for the way one should quantify
the efficiency of mixing in turbulent stratified fluids, with new
definitions for the mixing efficiency (or more accurately dissipations ratio)
$\gamma_{mixing}$ and flux 
Richardson number $R_f$ being proposed in the introduction, where they
are also compared with existing definitions.

\par

 A significant achievement of the extended APE framework is to allow for
a more rigorous and general re-derivation of MW98's result
(our Eq. (\ref{general_constraint})) that is also valid for a 
non-Boussinesq ocean, and which results in the appearance of a 
nonlinearity parameter $\xi \le 1$, with MW98's results being recovered 
for $\xi=1$. The main new result here is the finding that the work rate done
by the surface buoyancy fluxes $G(APE)$ should be numerically comparable
to $W_{r,forcing}$. This is important, because while $W_{r,forcing}$ is
currently widely agreed to be large $O(0.4\,{\rm TW})$, $G(APE)$ has been
widely thought to be negligible on the basis of MW98's argument that
this is required by Sandstrom's ``theorem''. The result therefore demonstrates
that $G(APE)$ is as important as the mechanical forcing in driving and
stirring the oceans, in agreement with \cite{Oort1994}'s previous conclusions.
The two main consequences are that: 1) there is no reason to reject the idea
that the oceans are a heat engine, and that the North-South heat transport
is mostly the result of the buoyancy forcing, as was usually thought to be
the case prior to MW98's study, \eg \cite{ACDV1993}; 2) the overall value
of $\gamma_{mixing}$ in the oceans is likely to be significantly larger
than the value of $\gamma_{mixing}=0.2$ currently used, as seems to be required
by the large magnitude of $G(APE)$, given that buoyancy-driven turbulent
mixing has a significantly higher $\gamma_{mixing}$ than mechanically-driven
turbulent mixing in general. Note that increasing $\gamma_{mixing}$ 
decreases the requirement on the mechanical sources of stirring, thereby
providing a natural way to remove 
the apparent paradoxical result of a shortfall
in the mechanical stirring energy that arises when 
$\gamma_{mixing}=0.2$ is used in Eq. (\ref{general_constraint}) for $\xi=1$.
 The extended APE framework may also be used to revisit the assumed 
implications of \cite{Paparella2002}'s anti-turbulence ``theorem'', the main
conclusion being that horizontal convection may in fact support elevated
values of diapycnal mixing, in contrast to what is usually assumed.
It seems important to stress, however, that even if one could prove 
that surface buoyancy fluxes could sustain on their own
diapycnal mixing rates as large as observed,
it would not prove that the actual buoyancy-driven
component of the ocean circulation is not mechanically-controlled. Indeed,
it is essential to recognise that both $G(KE)$ and $G(APE)$ do not depend
only on the external forcing parameters, but also strongly depend
on the actual ocean circulation and stratification, which necessarily implies
a mechanical control of the buoyancy-driven circulation and vice versa.
In other words, even though the present study disagrees with MW98's contention
that surface buoyancy fluxes do not work or stir the oceans, there should be
little doubt that even if the MOC is expected to be primarily buoyancy-driven,
it is also mechanically controlled in ways
that are not fully understood. In this respect, therefore, MW98 deserves
credit for challenging the idea that the behaviour of the buoyancy-driven
circulation can be understood independently of its link with the mechanical
forcing.

\par

An important challenge ahead 
will be to extend the present framework to deal
with an equation of state also depending on salinity, which gives rise to
the possibility of double-diffusion effects, and storing energy in chemical
form.
Such extensions are needed to better connect the present results to
many laboratory experiments based on the use of compositionally-stratified
fluids, although the present evidence is that the use of a linear equation
of state is probably accurate enough to describe the latter. To that end,
many technical difficulties need to be overcome. 
 Indeed, salinity complicates the definition of 
\cite{Lorenz1955}'s reference state to such an extent that it is not even
clear that such a state can be uniquely defined, \eg \cite{Huang2005}, as
may be the case in presence of humidity in the atmosphere, \eg
\cite{Tailleux2004}. A potentially important generalisation would also be
to further decompose the internal energy in order to isolate the available
acoustic energy considered by \cite{Bannon2004}, which in the present paper
is included as part of our definition of $APE$.
Finally, further work is required to understand
how to fix the value of $\xi$ and $\gamma_{mixing}$ in Eq. 
(\ref{general_constraint}) in the actual oceans.

\begin{acknowledgments}
  The author gratefully acknowledges discussions, comments, and 
a number of 
contrary opinions from J.C. McWilliams, J. Moelemaker, W. R. Young, 
T.J. McDougall, R.X. Huang, C. Staquet, G. Nurser, D. Webb, P. Muller,
A. Wirth, D. Straub, M. E. McIntyre, M. Ambaum, J. Whitehead,
T. Kuhlbrodt, R. Ferrari, D.P. Marshall, R. M. Samelson, R. A. deSzoeke,
G. Vallis, H.L. Johnson, A. Gnanadesikan, J. Nycander, O. Marchal,
J. Gregory, L. Rouleau and A. van der Linden,
which contributed to a much improved version of
two earlier drafts. I am particularly indebted to D. Straub and A. Wirth for
helping me realize that what I formerly interpreted as an artifact
of the Boussinesq approximation was in fact an artifact of using
a linear equation of state. I also thank
three anonymous referees for their helpful and constructive comments
which contributed enormously in improving the present manuscript.
As this paper was under review, 
G.O. Hughes, A. McC. Hogg, and R. W. Griffiths sent me
a manuscript illustrating nicely some of the points developed in the
present manuscript (\cite{Hughes2009}), for which I thank them.
 This study was
supported by the NERC funded RAPID programme.
\end{acknowledgments}

\appendix

\section{Energetics of Incompressible Navier-Stokes Equations}
\label{boussinesq_models}

\subsection{Boussinesq equations with equation of state nonlinear in
temperature}

 The purpose of this appendix is to document the energetics of
the Boussinesq system of equations that form the basis for most
inferences about stratified turbulence for fluid flows at low Mach
numbers, and which is commonly used in the theoretical
and numerical study of turbulence, \eg 
\cite{Winters1995,Caulfield2000,Staquet2000,Peltier2003}.
In order to go beyond the usual case of a linear equation of state,
a slight generalisation is introduced by allowing the thermal expansion
coefficient to vary with temperature.
The resulting set of equations is therefore as follows:
\begin{equation}
     \frac{D{\bf v}}{Dt} + \frac{1}{\rho_0}\nabla P = 
     - \frac{g \rho}{\rho_0} \hat{\bf z} + \nu \nabla^2 {\bf v}
\end{equation}
\begin{equation}
     \nabla \cdot {\bf v} = 0
\end{equation}
\begin{equation}
     \frac{DT}{Dt} = \kappa \nabla^2 T
     \label{temperature_equation}
\end{equation}
\begin{equation}
     \rho(T) = \rho_0 \left [ 1 - \int_{T_0}^{T} \alpha (T') dT' \right ]
     \label{eos_boussinesq}
\end{equation}
where ${\bf v}=(u,v,w)$ is the three-dimensional velocity
field, $P$ is the pressure, $\rho$ the density, $T$ the temperature,
$\nu=\mu/\rho$ the kinematic viscosity, $\mu$ the (dynamic) viscosity,
$\kappa$ the molecular diffusivity,
$g$ the acceleration of gravity, and $\rho_0$ a reference
density. The classical Boussinesq model, called L-Boussinesq model in this
paper, is simply recovered by taking $\alpha$ to be a constant in
Eq. (\ref{eos_boussinesq}). In that case, Eqs.
(\ref{temperature_equation}) and (\ref{eos_boussinesq}) may be combined
to obtain the
following diffusive model for density:
\begin{equation}
      \frac{D\rho}{Dt} = \kappa \nabla^2 \rho
\end{equation}
as assumed in many numerical studies of turbulence, \eg 
\cite{Winters1995,Caulfield2000,Staquet2000,Peltier2003}.
When the temperature dependence of $\alpha$ is retained,
the resulting model is called here the NL-Boussinesq model.

\subsection{Standard energetics}
 
  Evolution equations for the KE and GPE are obtained by the standard
procedure, \eg \cite{Batchelor1967,Landau1987}, assuming that the 
system is forced mechanically by an external stress ${\bf \tau}$, 
and thermodynamically by external heat fluxes, both assumed to act
at the surface boundary located at $z=0$.
The first equation is premultiplied by $\rho_0 {\bf v}$ and
volume-integrated. After re-organisation, the equation becomes:
\begin{equation}
   \frac{d(KE)}{dt} = 
   \frac{d}{dt} \int_{V} \rho_0 \frac{{\bf v}^2}{2} dV
   = \underbrace{\int_{\partial V} {\bf \tau}\cdot {\bf u}_s dS}_{G(KE)} 
   - \underbrace{\int_{V} \rho g w \, dV}_{W} 
   + \underbrace{\int_{V} \rho_0 \varepsilon \, dV}_{D(KE)}
   \label{KE_boussinesq}
\end{equation}
where $W$ is the so-called density flux, $D(KE)$ is the viscous dissipation
rate of kinetic energy, and $G(KE)$ is the rate of work done by the 
external stress. The time evolution of the total gravitational
potential energy of the fluid, i.e., the volume integral of $\rho g z$,
is:
\begin{equation}
    \frac{d(GPE)}{dt} = \frac{d}{dt} \int_{V} \rho g z \, dV
  = \underbrace{\int_{V} \rho g w \, dV}_{W}  
    \underbrace{-\int_{V} \rho_0 g z \alpha 
   \kappa \nabla^2 T \,dV}_{B} ,
   \label{GPE_boussinesq}
\end{equation}
where $B$ is the Boussinesq approximation of the work of 
expansion/contraction. In the present case, it is possible to derive
an explicit analytical formula for $B$:
$$
   B = - \int_{V} g z \rho_0 \alpha \kappa \nabla^2 T \,dV 
  =  \int_{V} \kappa g \rho_0 \alpha
      \frac{\partial T}{\partial z} \, dV
  + \int_{V} \kappa\rho_0 g z \frac{d\alpha}{dT}(T)
      \| \nabla T \|^2 \, dV
$$
\begin{equation}
     = \underbrace{\kappa g \left [ \langle \rho \rangle_{bottom} -
   \langle \rho \rangle_{top} \right ]}_{B_L} 
   + \int_{V} \kappa \rho_0 g z \frac{d\alpha}{dT}(T)
     \| \nabla T \|^2 \,dV,
   \label{B_formula}
\end{equation}
by using integration by parts, and using the fact that the surface term
vanishes because the surface is by assumption located at $z=0$,
where  $\langle \rho \rangle_{bottom}$ and $\langle \rho \rangle_{top}$ 
denote the surface-integral of the bottom and top value of density.
For a linear equation of state, $B=B_L$ will in general be small,
because of the smallness
of the molecular diffusivity $\kappa$, and finite top-bottom
density difference. When $\alpha$ increases with temperature, however,
$B$ may become significantly larger than $B_L$ in turbulence strong enough
as to make $\| \nabla T \|^2$ large enough for the second term in Eq. 
(\ref{B_formula}) to overcome $B_L$, pointing out the possibly critical
role of nonlinearity of the equation of state in strongly turbulent fluids.

\subsection{\cite{Lorenz1955}'s available energetics}

 We now seek evolution equations for the available and un-available parts
of the gravitational potential energy, as previously done by 
\cite{Winters1995} in the case of the L-Boussinesq equations. By definition,
the expression for the $GPE_r$ is:
\begin{equation}
    GPE_r = \int_{V} \rho_r g z_r dV ,
    \label{gper_appendix}
\end{equation}
where $z_r = z_r ({\bf x},t)$ and 
$\rho_r=\rho_r(z_r,t)$ are the vertical position
and density of the parcels in the \cite{Lorenz1955}'s reference state. 
In Boussinesq models, fluid parcels are assumed to conserve their
in-situ temperature in the reference state, so that
$T_r(z_r,t) = T ({\bf x},t)$. Taking the time derivative
of Eq. (\ref{gper_appendix}) thus yields: 
$$
   \frac{d(GPE_r)}{dt} = \int_{V} g z_r \frac{D\rho_r}{Dt} dV
     + \underbrace{\int_{V} g \rho_r \frac{Dz_r}{Dt} dV}_{=0}
    = - \int_{V} g z_r \rho_0 \alpha \kappa \nabla^2 T dV
$$
$$
  = -\int_{\partial V} g z_r \rho_0 \alpha \kappa \nabla T \cdot {\bf n}\,dS
  + \int_{V} \kappa \rho_0 g \nabla T \cdot \nabla ( \alpha z_r )  dV
$$
\begin{equation}
 =  - W_{r,forcing} +  
    \underbrace{\int_{V} \kappa \rho_0 g \| \nabla z_r \|^2 
    \frac{\partial (\alpha z_r)}{\partial z_r}
   \frac{\partial T_r}{\partial z_r} dV}_{W_{r,mixing}} ,
\end{equation}
where $W_{r,forcing}$ is the rate of change of $GPE_r$ due to the
external surface heating/cooling. For a Boussinesq fluid, this term
is identical to the $APE$ production rate $G(APE)$, as shown in the
following, i.e., $W_{r,forcing}=G(APE)$. The above formula was obtained
by using the following intermediate results:
\begin{equation}
   \nabla [T({\bf x})] = \nabla [T_r(z_r({\bf x}))] = 
     \frac{\partial T_r}{\partial z_r} \nabla z_r ,
\end{equation}
\begin{equation}
   \nabla [\alpha (T) z_r] = \nabla [ \alpha (T_r(z_r)) z_r ]
   = \frac{\partial (\alpha z_r)}{\partial z_r} \nabla z_r ,
\end{equation}
as well as the important result that the integral involving the
term $Dz_r/Dt$ vanishes identically established by \cite{Winters1995}.
In this paper, the result was established by using an explicit
formula for the reference stratification. An alternative way to 
recover such a result is achieved by noting that the velocity 
${\bf v}_r = (Dx_r/Dt,Dz_r/Dt)$ of the fluid 
parcels in the reference state must satisfy the continuity equation:
\begin{equation}
    \nabla_r \cdot {\bf v}_r = 0 ,
\end{equation}
where $\nabla_r$ is the divergence operator in the reference space
state $(x_r,z_r)$, from which it follows that the surface-integral
of $W_{r,mixing}=Dz_r/Dt$ along each constant $z_r$ level must vanish, which
implies \cite{Winters1995}'s result. The equation for $APE=AGPE
= GPE-GPE_r$ becomes:
$$
  \frac{d(APE)}{dt} = \frac{d(GPE)}{dt} - \frac{d(GPE_r)}{dt} 
$$
\begin{equation}
 = W - (W_{r,mixing}-B) + W_{r,forcing} = G(APE) + W - D(APE) ,
\end{equation}
where
$$
  D(APE) = W_{r,mixing} -B 
$$
\begin{equation}
    = \underbrace{\int_{V} \rho_0 g \alpha \kappa \| \nabla z_r \|^2 
      \frac{\partial T_r}{\partial z_r} dV - B_L}_{D_L(APE)} 
   + \underbrace{\int_{V} \kappa \rho_0 g (z_r-z) \frac{d\alpha}{dT} 
      \| \nabla z_r \|^2 \left ( \frac{\partial T_r}{\partial z_r}
     \right )^2 dV}_{D_{NL}(APE)} ,
\end{equation}
by using the results that:
\begin{equation}
      \int_{V} \rho_0 g \kappa \frac{d\alpha}{dT} \| \nabla T \|^2 dV
   = \int_{V} \rho_0 g \kappa \frac{d\alpha_r}{dT_r} \| \nabla z_r \|^2
     \left (\frac{\partial T_r}{\partial z_r} \right )^2 dV ,
\end{equation}
\begin{equation}
     \int_V \rho_0 g \kappa \| \nabla z_r \|^2 
     \frac{\partial (\alpha z_r)}{\partial z_r} 
     \frac{\partial T_r}{\partial z_r} dV
    = \int_{V} \rho_0 g \kappa \| \nabla z_r \|^2 \left ( 
     1 + z_r \frac{d\alpha}{dT} \frac{\partial T_r}{\partial z_r}
     \right ) \frac{\partial T_r}{\partial z_r} dV .
\end{equation}
Empirically, it is usually found that $D(APE)>0$, which is not readily
apparent from the form of $D(APE)$, and for which a rigorous mathematical
proof remains to be established. Interestingly, while $W_{r,mixing}$ and
$B$ appear to be both strongly modified by a temperature-dependent $\alpha$,
this is much less so for their difference $D(APE)$, which is usually found
empirically to be well approximated by its ``linear'' part $D_L(APE)$.
This is important, because it clearly establishes that $D(APE)$ and
$W_{r,mixing}$ may be significantly different when the temperature 
dependence of $\alpha$ is retained, in contrast to what is generally
admitted based on the L-Boussinesq model. This suggests that results
based on the study of the L-Boussinesq model are likely to be more robust
and accurate
for the description of the $KE/APE$ dynamics than for the description of
$GPE_r$.
The condition for $|D_{NL}(APE)| \ll |D_L(APE)|$ to be satisfied is that 
$d\alpha/dT|dT_r/dz_r||z_r-z| \ll 1$, which appears to be satisfied in
practice for water or seawater. Whether this is also true for other types
of fluids still needs to be established.

\section{Energetics of compressible Navier-Stokes Equations}
\label{energetics_cnse}

\subsection{Compressible Navier-Stokes Equations (CNSE)}

  The purpose of this appendix is to generalise \cite{Winters1995}'s results
to the fully compressible Navier-Stokes equations, which are written here
in the following form:
\begin{equation}
    \rho \frac{D{\bf v}}{Dt} + \nabla P = -\rho g \hat{\bf z} +
     \nabla \cdot {\bf S}
    \label{momentum_equation}
\end{equation}
\begin{equation}
     \frac{D\rho}{Dt} + \rho \nabla \cdot {\bf v} = 0
     \label{mass_equation}
\end{equation}
\begin{equation}
     \frac{D\Sigma}{Dt} = \frac{\dot{Q}}{T}=
     \frac{\rho \varepsilon - \nabla \cdot {\bf F}_q}{\rho T}
     \label{entropy_equation}
\end{equation}
\begin{equation}
       I = I(\Sigma, \upsilon),
      \label{ie_eos}
\end{equation}
\begin{equation}
      T = T(\Sigma,\upsilon) = \frac{\partial I}{\partial \Sigma}, \qquad
      P = P(\Sigma,\upsilon) = - \frac{\partial I}{\partial \upsilon}
      \label{tp_eos} .
\end{equation}
In the present description,  
the three-dimensional Eulerian velocity field ${\bf v}=(u,v,w)$,
the specific volume $\upsilon=1/\rho$ (with $\rho$ the density),
and the specific entropy $\Sigma$ are taken as the dependent variables,
with the thermodynamic pressure $P$ and in-situ temperature $T$ 
being diagnostic variables  
 as expressed by 
Eqs. (\ref{ie_eos}-\ref{tp_eos}), where $I$ is the specific internal
energy, regarded as a function of $\Sigma$ and $\upsilon$.
Further useful notations are:
$D/Dt = \partial/\partial t  + ({\bf v}\cdot \nabla )$ is the substantial
derivative, $\varepsilon$ is the dissipation rate of kinetic 
energy, ${\bf F}_q = -k_T \rho C_p \nabla T$ is the diffusive heat flux, 
$C_p$ is the specific heat capacity at constant pressure,
$k_T$ is the molecular diffusivity for temperature, $g$ is the acceleration
of gravity, and ${\bf z}$ is a
normal unit vector pointing upward. Moreover, ${\bf S}$ is the deviatoric
stress tensor:
\begin{equation}
     S_{ij} = \mu \left ( \frac{\partial u_i}{\partial x_j}
  + \frac{\partial u_j}{\partial x_i} \right )  + 
     \left ( \lambda - \frac{2\mu}{3} \right )
    \delta_{ij} \frac{\partial u_{\ell}}{\partial x_{\ell}}
\end{equation}
in classical tensorial notation, e.g., \cite{Landau1987},
where Einstein's summation convention
for repeated indices has been adopted, and where $\delta_{i,j}$ is
the Kronecker delta. The parameters $\mu$ and $\lambda$ are the
shear and bulk (or volume) viscosity respectively.

\subsection{Standard energetics}

  The derivation of evolution equations for the standard forms of energy
in the context of the fully compressible Navier-Stokes equations is a 
standard exercise, \eg \cite{degroot1962,Landau1987,Griffies2004}, 
so that only the final results are given. In the standard description
of energetics, only the volume-integrated kinetic energy (KE), 
gravitational potential energy (GPE), and internal energy (IE) are
considered, viz., 
\begin{equation}
   KE = \int_{V} \rho \frac{{\bf v}^2}{2}\, dV, \qquad
  GPE = \int_{V} \rho g z dV\, \qquad
  IE = \int_{V} \rho I(\Sigma,\upsilon)\,dV ,
\end{equation}
whose standard evolution equations are respectively given by:
\begin{equation}
   \frac{d(KE)}{dt} = -\underbrace{\int_{V} \rho g w\, dV}_{W} + 
   \underbrace{\int_{V}
   P \frac{D\upsilon}{Dt}}_{B} \, dm + G(KE) - D(KE) 
   -P_a \frac{dV_{ol}}{dt} ,
  \label{KE_equation}  
\end{equation}
\begin{equation}
    \frac{d(GPE)}{dt} = \underbrace{\int_{V} \rho g w \,dV}_{W} ,
  \label{GPE_equation}
\end{equation}
\begin{equation}
   \frac{d(IE)}{dt} = \int_{V} \rho \dot{Q}\,dV - 
   \underbrace{\int_{V} P \frac{D\upsilon}{Dt}\, dm }_{B} 
   = D(KE) + Q_{heating} - Q_{cooling} - B ,
  \label{IE_equation}
\end{equation}
where $G(KE)$ is the rate of work done by the mechanical sources of energy
on the fluid,
$Q_{heating}$ (resp. $Q_{cooling})$ is the surface-integrated rate of 
heating  (resp. cooling) due to the thermodynamic sources of energy,
and $V_{ol}$ is the total volume of the fluid; additional definitions and 
justifications are given further down the text. Summing Eqs. 
(\ref{KE_equation}-\ref{IE_equation}) yields the following evolution 
equation for the total energy total energy $TE=KE+GPE+IE$:
\begin{equation}
   \frac{d(TE)}{dt} = G(KE) + Q_{heating}- Q_{cooling}
      - P_a \frac{dV_{ol}}{dt} ,
\end{equation}
which says that the total energy of the fluid is modified:
\begin{itemize}
\item by the rate of work done by the mechanical sources of energy;
\item by the rate of heating/cooling 
done by the thermodynamic sources of
energy;
\item by the rate of work done by the atmospheric pressure $P_a$ against the
volume changes of the fluid.
\end{itemize}
 As these derivations are quite
standard, justifications for the above equations are only briefly outlined.
Thus, the KE equation (\ref{KE_equation}) is
classically obtained by multiplying the momentum equation by ${\bf v}$,
and integrating over the volume domain. The term $W$ results from the
product of ${\bf v}$ with the gravitational force vector, whereas 
the product ${\bf v}\cdot \nabla P = \nabla \cdot (P{\bf v})
- P \nabla \cdot {\bf v} = \nabla \cdot (P{\bf v}) - (P/\upsilon)
D\upsilon/Dt$ yields the work of expansion/contraction minus the
work done by the atmospheric pressure against total volume changes.
The product of the velocity vector with the stress tensor is written
as the sum $G(KE)-D(KE)$, where $G(KE)$ represents the work input
due to the external stress, and $D(KE)$ the positive dissipation of kinetic
energy. The general expression for the mechanical energy input is:
\begin{equation}
      G(KE) = \int_{\partial V} {\bf v}{\bf S} \cdot {\bf n} dS
      = \int_{\partial V} {\bf \tau} \cdot {\bf v} dS 
\end{equation}
where ${\bf v}{\bf S}$ is the vector with components $({\bf S}{\bf v})_j
= S_{ij}u_i$, while ${\bf S}{\bf n}={\bf \tau}$ is the stress applied
along the surface boundary enclosing the fluid. $G(KE)$ is therefore
the work of the applied stress done against the fluid velocity. If one
assumes no-slip boundary condition on all solid boundaries, then this
work is different from zero only on the free surface.
The function $D(KE)$ is the dissipation function:
\begin{equation}
      D(KE) = \int_{V} \left \{ \mu \left ( \frac{\partial u_i}{\partial x_j}
    + \frac{\partial u_j}{\partial x_i} - \frac{2}{3} 
      \delta_{ij} \frac{\partial u_{\ell}}{\partial x_{\ell}}
      \right )^2 + \lambda ( \nabla \cdot {\bf v} )^2 \right \} dV ,
\end{equation}   
where again the summation convention for repeated indices has been used,
e.g., \cite{Landau1987}.
 The equation for $GPE$ (\ref{GPE_equation})
is simply obtained by taking the time derivative of its definition,
using the fact that $D(\rho g z dV)/Dt = \rho g w$, since
$D(\rho dV)/Dt=0$ from mass conservation. The equation for $IE$
(\ref{IE_equation}) results from the fact that the differential of
internal energy in the entropy/specific volume representation is
given by $dI=Td\Sigma - Pd\upsilon$. The terms $Q_{heating}$ and
$Q_{cooling}$ represents the surface-integrated net heating and
cooling respectively going through the surface enclosing the domain.

\subsection{Available energetics}

 In this paragraph, we seek to derive separate evolution equations for
the available and un-available parts of the total potential energy
$PE=IE+GPE+P_a V_{ol}$, as initially proposed by \cite{Lorenz1955}, 
building upon ideas going back to \cite{Margules1903}. Specifically,
$PE$ is decomposed as follows:
$$
  PE = \int_{V} \rho \left [ I(\Sigma,\upsilon) + g z \right ]\,dV + 
   P_a V_{ol}
$$
$$
  = \underbrace{\int_{V} \rho \left [ I(\Sigma,\upsilon) + g z \right ]\,dV
  - \int_{V} \rho \left [ I(\Sigma,\upsilon_r) + g z_r \right ]\,dV
 + P_a \left ( V_{ol} - V_{ol,r} \right )}_{APE}
$$
\begin{equation}
  + \underbrace{ 
 \int_{V} \rho \left [ I (\Sigma,\upsilon_r) + g z_r \right ]\,dV
   + P_a V_{ol,r}}_{PE_r}
  \label{pe_decomposition}
\end{equation}
where $PE_r$ is the potential energy of \cite{Lorenz1955}'s reference state,
and $APE=PE-PE_r$ is the available potential energy.
As is well known, the reference state is the state minimising the total
potential energy of the system in an adiabatic re-arrangement of the fluid
parcels. From a mathematical viewpoint, \cite{Lorenz1955}'s reference state
can be defined in terms of a mapping taking a
parcel located at $({\bf x},t)$ in the given
state to its position $({\bf x}_r,t)$ in the reference state, such 
that the mapping 
preserves the specific entropy $\Sigma$ and mass $\rho dV$ of the parcel,
viz.,
\begin{equation}
     \Sigma ({\bf x},t) = \Sigma ({\bf x}_r,t) = \Sigma_r (z_r,t) ,
\end{equation}
\begin{equation}
     \rho ({\bf x},t) dV = \rho({\bf x}_r, t) dV_r =
     \rho_r (z_r,t) dV_r ,
\end{equation}
where the second condition can be equivalently formulated in terms of
the Jacobian $J=\partial ({\bf x}_r)/\partial ({\bf x})$ of the
mapping between the actual and reference state as follows:
\begin{equation}
     \rho ({\bf x},t) = \rho({\bf x}_r,t) 
      \frac{\partial ({\bf x}_r)}{\partial ({\bf x})} =
     \rho_r (z_r,t) \frac{\partial ({\bf x}_r)}{\partial ({\bf x})} .
\end{equation}
Prior to deriving evolution equation for $PE_r$ and $APE$, it is  
useful to mention three important properties of the reference state, namely:
\begin{enumerate}
\item The density $\rho_r = \rho_r (z_r,t)$ and pressure 
$P_r = P_r(z_r,t)$ of the \cite{Lorenz1955}'s background reference state
are functions of $z_r$ alone (and time);
\item The background density $\rho_r$ and pressure $P_r$ are in hydrostatic
balance at all times, i.e.,  $\partial P_r/\partial z_r = -\rho_r g$ 
(this is a consequence of the reference state
being the state minimising the total potential energy in an adiabatic
re-arrangement of the parcels);
\item The velocity 
${\bf v}_r = (Dx_r/Dt, Dy_r/Dt, Dz_r/Dt)$ of the parcels in the reference
state satisfies the usual mass conservation equation:
\begin{equation}
    \frac{D\upsilon_r}{Dt} = \upsilon_r \nabla_r \cdot {\bf v}_r ,
    \label{mass_conservation_r}
\end{equation}
where $\nabla_r \cdot {\bf v}_r$ is the velocity divergence expressed
coordinates system of the reference state,
which is a consequence of the mass of the fluid parcels being conserved
by the mapping between the actual and reference state.
\end{enumerate}
Eq. (\ref{mass_conservation_r}) is important, for it allows an easy
demonstration of the following result:
$$
  \int_{V} \rho P_r \frac{D\upsilon_r}{Dt} dV =
  \int_{V_r} \rho_r P_r \frac{D\upsilon_r}{Dt} dV_r
 = \int_{V_r} P_r \nabla_r \cdot {\bf v}_r dV_r 
$$
\begin{equation}
  = \int_{\partial V_r} P_r {\bf v}_r \cdot {\bf n}_r dS_r
   - \int_{V_r} {\bf v}_r \cdot \nabla P_r dV_r
  = P_a \frac{dV_{ol,r}}{dt} + 
   \underbrace{\int_{V_r} \rho_r g w_r dV_r}_{W_r} ,
  \label{pdv_r}
\end{equation}
which establishes the equivalence between the work of expansion and
the work against gravity in the reference state,
where ${\bf n}_r$ is a outward pointing unit vector normal to the
boundary $\partial V_r$ enclosing the fluid in the reference state.
In Eq. (\ref{pdv_r}), the first equality stems from expressing the
first integral in the reference state; the second equality uses
Eq. (\ref{mass_conservation_r}); the third equality results from
an integration by parts; the final equality stems from that $P_r$
depends on $z_r$ and $t$ only, and that it is in hydrostatic balance,
and from using the boundary condition ${\bf v}_r\cdot {\bf n}_r 
= w_r = \partial \eta_r/\partial t$ at the surface assumed to be
located at $z_r=\eta_r(t)$.

\subsection{Evolution of the background potential energy $PE_r$}

  We seek an evolution equation for the background $PE_r$ by taking
the time derivative of the expression in Eq. (\ref{pe_decomposition}),
which yields:
$$
  \frac{d(PE_r)}{dt} = \int_{V} \rho \left [ 
   T_r \frac{D\Sigma}{Dt} - P_r \frac{D\upsilon_r}{Dt} 
  + g w_r \right ]\,dV + P_a \frac{dV_{ol,r}}{dt}
  + \underbrace{\int_{V} \left [ I(\Sigma,\upsilon_r)+g z_r \right ]
   \frac{D(\rho dV)}{Dt}}_{=0}
$$
$$
   = \int_{V} \rho T_r \frac{\dot{Q}}{T} \,dV
  =  \int_{V} \rho \dot{Q}\,dV + \int_{V} \rho 
   \left ( \frac{T_r-T}{T} \right )\dot{Q} \,dV 
$$
\begin{equation}
    = \dot{Q}_{net} + (1-\gamma_{\varepsilon}) D(KE) + D(APE) - G(APE) ,
\end{equation}
where the final result was arrived at by making use of Eq. (\ref{pdv_r}),
as well as of the definitions:
\begin{equation}
    \int_{V} \rho \dot{Q} \,dV = \int_{V}\left \{ \nabla \cdot \left (
   \kappa \rho C_p \nabla T \right ) + \rho \varepsilon \right \}\,dV =
   \dot{Q}_{net} + D(KE),
\end{equation}
\begin{equation}
      \dot{Q}_{net} = \int_{S} \kappa \rho C_p \nabla T \cdot {\bf n}\,dS
  =  Q_{heating} - Q_{cooling}
   \label{Q_net}
\end{equation}
\begin{equation}
      G(APE) = \int_{S} \left ( \frac{T-T_r}{T} \right ) 
   \kappa \rho C_p \nabla T \cdot {\bf n} \,dS
   \label{gape_definition}
\end{equation}
\begin{equation}
     D(APE) =  \int_{V} \kappa \rho C_p \nabla T \cdot
     \nabla \left ( \frac{T-T_r}{T} \right ) \,dV,
     \label{dape_definition}
\end{equation}
\begin{equation}
   \gamma_{\varepsilon} D(KE) =
   \int_{V} \left ( \frac{T - T_r}{T} \right ) \rho \varepsilon \, dV
   \label{gamma_epsilon}
\end{equation} 
where ${\bf n}$ is the unit normal vector pointing outward the domain.
Eq. (\ref{Q_net}) expresses the net diabatic heating $\dot{Q}_{net}$ due
to the surface heat fluxes as the sum of a purely positive $Q_{heating}$
and negative $-Q_{cooling}$ contributions.
Eq. (\ref{gape_definition}) defines the rate of available potential energy
produced by the surface heat fluxes. The term $D(APE)$, as defined by
Eq. (\ref{dape_definition}), is physically expected to represent the rate
at which $APE$ is dissipated by molecular diffusion, so that it is expected
to be positive in general, which has been so far only established
empirically using randomly generated temperature fields, but a rigorous
mathematical proof is lacking. Finally, 
Eq. (\ref{gamma_epsilon}) states that a tiny fraction
of the diabatic heating due to viscous dissipation might be recycled
to produce work. If $\gamma_{\varepsilon}$ could be proven to be positive,
it could probably be included as part of the $G(APE)$. In the following,
it will just be neglected for simplicity.

\subsection{Evolution of Available Potential Energy (APE)}

\begin{figure}
\centering
\includegraphics[width=13cm]{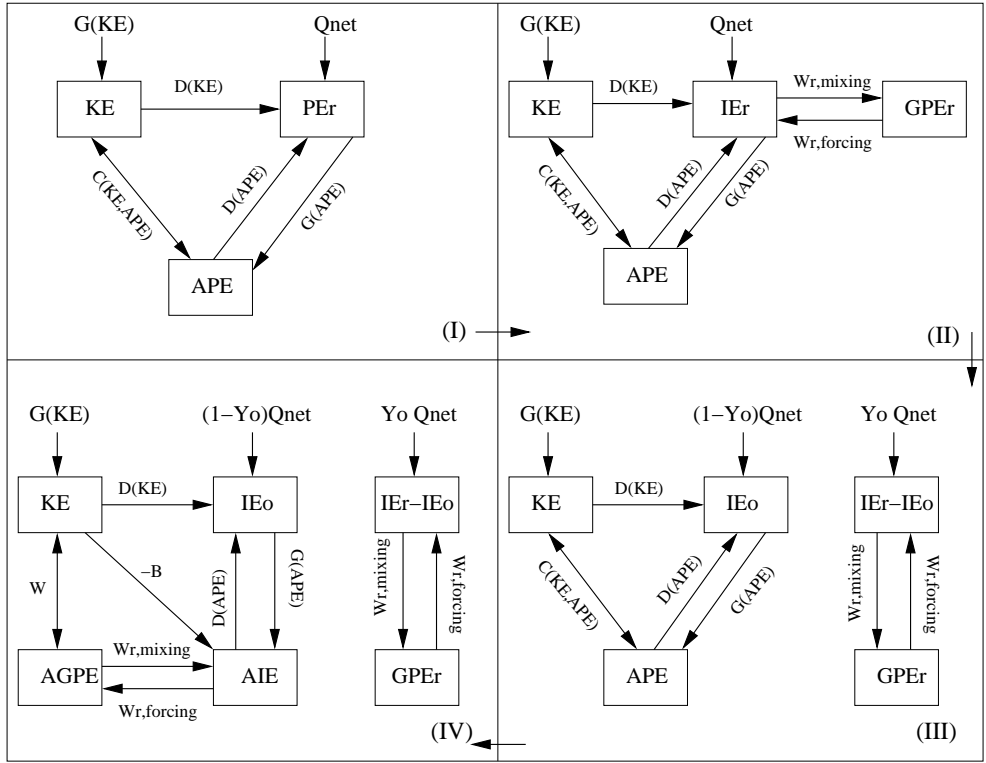}
\caption{Successive refinements of the energetics of a forced/dissipated
stratified fluid. Panel (I): $KE/APE/PE_r$ representation. Panel (II):
Decomposition of $PE_r$ into $IE_r+GPE_r$. Panel (III): Decomposition
of $IE_r$ into a dead part $IE_0$ and exergy part $IE_{exergy}=IE_r-IE_0$.
Panel (IV): Decomposition of $APE$ into $AIE$ and $AGPE$, revealing the
link between $C(KE,APE)$ to the density flux $W$ and work of 
expansion/contraction $B$.}
\label{basic_energetics}
\end{figure}

In the previous section, we defined the total potential energy as the
sum of $GPE$, $IE$, and the quantity $P_a V_{ol}$, see 
Eq. (\ref{pe_decomposition}). As a result, using the evolution equations
for $GPE$ and $IE$ previously derived, the evolution equation for $PE$
is given by:
\begin{equation}
   \frac{d(PE)}{dt} = W - B + \dot{Q}_{net} + P_a \frac{dV_{ol}}{dt} .
\end{equation}
Now, combining this equation with the one previously derived for $PE_r$
allows us to derive the following equation for the available potential
energy $APE=PE-PE_r$, defined as the
difference between the potential energy and its background value:
$$
    \frac{d(APE)}{dt} \approx \underbrace{W - B + 
    P_a \frac{dV_{ol}}{dt}}_{C(KE,APE)} +
   \dot{Q}_{net} + D(KE) 
   - \left [D(KE) + D(APE) + \dot{Q}_{net}
      - G(APE) \right ] 
$$
\begin{equation} 
     = C(KE,APE) +  G(APE) - D(APE),
\end{equation}
where the final expression neglects the small term $\gamma_{\varepsilon}
D(KE)$.
The corresponding energy flowchart for the $KE/APE/PE_r$ system is very
simple, and is illustrated in Fig. \ref{basic_energetics} (Panel I). 
This diagram shows that mechanical energy enters the fluid via the $KE$
reservoir, and that thermal energy enters it via the $PE_r$ reservoir.
There are two dissipation routes associated with the viscous dissipation
of $KE$ and the diffusive dissipation of $APE$. Only a certain part
$G(APE)$ of the thermodynamic energy input can be converted into $APE$
and hence into $KE$, which is processed via the $PE_r$ reservoir.
The two-headed arrow indicates the reversible
conversion between $KE$ and $APE$.

\subsection{Splitting of $PE_r$ into $GPE_r$ and $IE_r$}

  Although the $KE/APE/PE_r$ system offers a simple picture of the 
energetics of a (turbulent or not) stratified fluid, it is useful to
further decompose the background $PE_r$ reservoir into its $GPE$ and
$IE$ component, in order to establish the link with the existing literature
about turbulent mixing, as well as with \cite{Munk1998}'s theory.
The particular question to be addressed is to understand how much of
$D(KE)$ and $D(APE)$ are actually spread over $GPE_r$ and $IE_r$.
Likewise, to what extent do $\dot{Q}_{net}$ and $G(APE)$ affect 
$GPE_r$ compared to $IE_r$, where we have the following definitions:
\begin{equation}
    IE_r = \int_{V} \rho({\bf x},t) I(\Sigma,\upsilon_r) \,dV = 
    \int_{V_r} \rho_r(z_r,t) I(\Sigma,\upsilon_r) \,dV_r
    \label{IEr_definition_bis}
\end{equation}
\begin{equation}
    GPE_r = \int_{V} \rho({\bf x},t) g z_r ({\bf x},t) dV =
            \int_{V_r} \rho_r(z_r,t) g z_r dV_r ,
    \label{GPEr_definition}
\end{equation}
by expressing the integrals in the coordinate system associated either
with the actual state or reference state.
By definition, 
\begin{equation}
   \frac{d(GPE_r)}{dt} = 
   \int_{V} \rho g \frac{Dz_r}{Dt} \,dV +
    \underbrace{\int_{V} g z_r \frac{D(\rho dV)}{dt}}_{=0} 
   = \int_{V} \rho g w_r \,dV = W_r
\end{equation}
so that the evolution equation for $IE_r + P_a V_{ol,r}=PE_r - GPE_r$ 
is simply:
\begin{equation}
    \frac{d(IE_r+P_a V_{ol,r})}{dt}  = 
         \dot{Q}_{net} + D(KE) + D(APE) - G(APE) - W_r
\end{equation}
In order to make progress, we need to relate $W_r$ to the different
sources and sinks affecting $PE_r$, as identified in Fig. 
(\ref{basic_energetics}). To that end, we use the fact that $W_r$ is
related to the work of expansion in the reference state, as shown by
Eq. (\ref{pdv_r}), and regard 
$\upsilon=\upsilon (\Sigma,P)$ as a function of entropy and pressure,
for which the total differential is given by:
\begin{equation}
    d\upsilon = \Gamma d\Sigma - \frac{1}{\rho^2 c_s^2} dP
\end{equation}
where $\Gamma = \alpha T/(\rho C_p)$ is the so-called adiabatic lapse rate,
\eg \cite{Feistel2003}, and $c_s^2 = (\partial P/\partial \rho)_{\Sigma}$ 
is the
squared sound of speed. As a result, the expression for $W_r$ becomes:
\begin{equation}
   W_r = \int_{V_r} P_r \frac{D\upsilon_r}{Dt} \rho_r\,dV_r
      - P_a \frac{dV_{ol,r}}{dt}
      = \int_{V_r} P_r' \rho_r \left [
        \frac{\alpha_r T_r}{\rho_r C_{pr}} \frac{\dot{Q}}{T} 
       - \frac{1}{\rho_r^2 c_{sr}^2} \frac{DP_r}{Dt} \right ] \,dV_r
   \label{wr_explicit}
\end{equation}
where $P_r' = P_r - P_a$ is the pressure corrected by the atmospheric
pressure, by noting that we have:
\begin{equation}
      \frac{dV_{ol,r}}{dt} = \int_{V} \frac{D \upsilon_r}{Dt} \rho\,
     dV = \int_{V_r} \frac{D\upsilon_r}{Dt}\,\rho_r dV_r .
\end{equation}
In order to simplify Eq. (\ref{wr_explicit}), let us recall that mass
conservation can be rewritten in hydrostatic pressure coordinates as
follows:
\begin{equation}
     \nabla_r \cdot {\bf u}_r + \frac{\partial}{\partial P_r}
    \frac{DP_r}{Dt} = 0
    \label{mass_p}
\end{equation}
\eg \cite{Haltiner1980,Deszoeke2002}. As a result, it follows that
integrating Eq. (\ref{mass_p}) from the surface where $P_a=cst$, and
hence where $DP_r/Dt=0$, to an arbitrary level indicates that 
the surface integral of $DP_r/Dt$ must vanish along any 
$z_r =cst$ surfaces. As a consequence, the term depending on $DP_r/Dt$ in
Eq. (\ref{wr_explicit}) must vanish. For an alternative derivation of this
result, see \cite{Pauluis2007}. The remaining term can be written
as follows:
$$
    W_r = \int_{V} \frac{P_r'\alpha_rT_r}{\rho_r C_{pr}T}
   \left \{ \nabla \cdot \left ( \kappa \rho C_p \nabla T \right )
  + \rho \varepsilon \right \} \,dV
$$
$$
  = \int_{V} \frac{P_r'\alpha_r}{\rho_r C_{pr}} \left (
    1 + \frac{T_r-T}{T} \right ) \nabla \cdot \left ( \kappa \rho C_p 
   \nabla T \right )\,dV + 
   \int_{V} \frac{P_r'\alpha_r T_r}{\rho_r C_{pr} T}\rho \varepsilon\,dV
$$
$$
     \underbrace{\int_{V} \frac{P_r'\alpha_r}{\rho_r C_{pr}} \nabla \cdot
    \left ( \kappa \rho C_p \nabla T \right )\,dV}_{W_{r,mixing}
    - W_{r,forcing}}
  + \underbrace{\int_{V} \frac{P_r'\alpha_r}{\rho_r C_{pr}} \left (
     \frac{T_r-T}{T} \right ) \nabla \cdot \left ( \kappa \rho C_p \nabla T
    \right )\,dV}_{\Upsilon_{r,ape}D(APE)}
    + \underbrace{\int_{V} \frac{P_r'\alpha_r T_r}{\rho_r C_{pr}T}
    \rho \varepsilon\,dV}_{\Upsilon_{r,ke}D(KE)}
$$
\begin{equation}
   = W_{r,mixing} - W_{r,forcing} + \Upsilon_{r,ape}D(APE)
   + \Upsilon_{r,ke} D(KE) 
  \label{wr_final}
\end{equation}
where we defined:
\begin{equation}
     W_{r,mixing} = - \int_{V} \kappa \rho C_p \nabla T \cdot \nabla 
    \left ( \frac{\alpha_r P_r'}{\rho_r C_{pr}} \right ) \,dV
\end{equation}
\begin{equation}
     W_{r,forcing}
     = -\int_S \frac{\alpha_r P_r'}{\rho_r C_{pr}} 
       \kappa \rho C_p \nabla T \cdot {\bf n} \,dS
     \label{wrforcing_definition}
\end{equation}
\begin{equation}
     \Upsilon_{r,ape} D(APE) = 
     \int_{V} \frac{\alpha_r P_r'}{\rho_r C_{pr}} \left (
      \frac{T_r-T}{T} \right ) \nabla \cdot \left ( 
      \kappa \rho C_p \nabla T \right )  dV
\end{equation}
\begin{equation}
     \Upsilon_{r,ke} D(KE) = 
     \int_{V} \frac{\alpha_r P_r}{\rho_r C_{pr}} \frac{T_r}{T} 
     \rho \varepsilon \,dV
\end{equation}
Eq. (\ref{wr_final}) shows that the variations of $GPE_r$ are affected by:
\begin{itemize}
\item Turbulent mixing, associated with $W_{r,mixing}$. This expression
is similar to the one previously derived for the $L$-Boussinesq model.
The classical Boussinesq expression can be recovered from using the
approximation $T\approx T_r$,
taking $\alpha_r$, $\rho_r$, and $C_{pr}$ as constant, and using the
approximation $P_r \approx = -\rho_0 g z_r$, which yields:
$$
   W_{r,mixing} \approx \int_{V} \kappa \rho_0 g \| \nabla z_r \|^2 
      \alpha \frac{\partial T_r}{\partial z_r} dV ;
$$

\item The surface forcing, associated with $W_{r,forcing}$. Likewise,
the $L$-Boussinesq expression can be recovered by making the same
approximation, yielding:
$$
    W_{r,forcing} \approx \int_{V} \frac{\alpha g z_r}{C_p} Q_{surf} dS
$$
Note that in the $L$-Boussinesq approximation, we have:
$$
    W_{r,forcing} \approx G(APE)
$$
which is not generally true in the fully compressible Navier-Stokes
equation;

\item The contribution from the viscous and diffusive dissipation of
$KE$ and $APE$ respectively associated with $D(KE)$ and $D(APE)$. Note
that the coefficient $\Upsilon_{r,ape}$ and $\Upsilon_{r,ke}$ are very
small for a nearly incompressible fluid such as seawater. For instance,
typical values are: $\alpha = 10^{-4}\,{\rm K}^{-1}$, $P=4.10^7\,{\rm Pa}$,
$\rho = 10^3 \,{\rm kg.m^{-3}}$, and $C_p = 4.10^3\,{\rm J.kg^{-1}.K^{-1}}$,
which yields:
$$
   \Upsilon_{r} = O \left ( \frac{10^{-4}\times 4.10^7}{10^3\times
    4.10^3} \right ) = O \left ( 10^{-3} \right ) .
$$
\end{itemize}
From this, it follows that at leading order, the direct effects of
$D(APE)$ and $D(KE)$ on $GPE_r$ can be safely neglected compared to
the other two effects, so that:
\begin{equation}
    \frac{d(GPE_r)}{dt} = W_r \approx W_{r,mixing} - W_{r,forcing}
\end{equation}
The resulting modifications to the energy flowchart are then displayed
in Fig. \ref{basic_energetics} (Panel II). At leading order, the 
effects of the forcing and mixing on $GPE_r$ appear as conversion terms
with $IE_r$.

\subsection{Further partitioning of internal energy into
a ``dead'' and ``exergy'' component}

  As seen previously, the L-Boussinesq model is such that:
\begin{equation}
         D(APE) \approx W_{r,mixing}, \qquad
          G(APE) \approx W_{r,forcing},
\end{equation}
which may give the impression, based on Fig. \ref{basic_energetics} 
(Panel II)
that the $APE$ dissipated by $D(APE)$ is actually converted into $GPE_r$,
while $G(APE)$ may also appear as originating from $GPE_r$. The purpose
of the following is to show that this is actually not the case. To that end,
we introduce an equivalent isothermal state having exactly the same energy
as \cite{Lorenz1955}'s reference state, i.e., that is defined by:
\begin{equation}
    \underbrace{IE_r + GPE_r + P_a V_{ol,r}}_{PE_r}
   = \underbrace{IE_0 + GPE_0 + P_a V_{ol,0}}_{PE_0} .
\end{equation}
Because both Lorenz's reference state and the equivalent thermodynamic
equilibrium state are in hydrostatic balance at all time, $PE_r$ and
$PE_0$ are just the total enthalpies of the two states. This makes
it possible to define each parcel by their horizontal coordinates $(x,y)$
and hydrostatic pressure $P$, and to assume that the dead state can be
obtained from Lorenz's reference state by an isobaric process, so that
$(x_0,y_0,P_0)=(x_r,y_r,P_r)$, which in turn implies
$(dx_0/dt,dy_0/dt,dP_0/dt)=(dx_r/dt, dy_r/dt,dP_r/dt)$. 

\par

  Prior to looking at the evolution of the dead state, let us establish
that if the pressure $P$ is in hydrostatic balance at all times, then
we have the following result:
\begin{equation}
      \int_{V} \frac{DP}{Dt} \,dV = \int_{V} {\bf u} \cdot \nabla_h P \,dV,
      \label{DPDT_equation}
\end{equation}
 where ${\bf u}$ is the
horizontal part of the 3D velocity field, and $\nabla_h$ the horizontal
nabla operator. The proof is:
$$
  \int_{V} \left ( \frac{DP}{Dt} - {\bf u} \cdot \nabla_h P \right )dV 
= \int_{V} \left ( 
   \frac{\partial P}{\partial t} + w \frac{\partial P}{\partial z} \right ) 
 dV = \frac{d}{dt} \int_{V} P\,dV - P_a \frac{dV_{ol}}{dt}
     - \frac{d}{dt} \int_{V} \rho g z dV
$$
$$
   = \frac{d}{dt} \left \{  P_a V_{ol} + M_{tot} g H_b  + 
  \int_{V} \rho g z dV \right \} - P_a \frac{dV_{ol}}{dt}
   - \frac{d}{dt} \int_{V} \rho g z dV = 0
$$
where $V_{ol}$ and $M_{tot}$ are the total volume and mass of the fluid,
whose expressions are:
$$
   V_{ol} = \int_{S} (\eta(x,y,t)+H_b )dxdy, \qquad
   g M_{tot} = \int_{S} (P_b(x,y,t)-P_a ) dx dy,
$$ 
where $z=\eta(x,y,t)$ is the equation for the free surface, $P_b(x,y,t)$
is the bottom pressure, $H_b$ is the total depth of the basin, and
where the expression between brackets was obtained by using the
following result:
$$
  \int_{V} P dV = \int_{S} [ P z ]_{-H_b}^{\eta} dx dy + 
  \int_{V} \rho g z dV = \int_{S} [P_a \eta + P_b H_b ]dx dy + \int_{V}
   \rho g z dV  
$$
$$
  = \underbrace{P_a \int_{S} (\eta+H)dxdy}_{P_a V_{ol}}
   + \underbrace{H \int_{S} (P_b-P_a) dx dy}_{M_{tot} g H} 
   + \int \rho g z dV . 
$$ 
 The important consequence of Eq. (\ref{DPDT_equation})
is that the volume integral of 
$DP/Dt$ identically vanishes when $P$ is independent of the horizontal
coordinates, as is the case for $P_r$ and $P_0$. Now, using the expression
for the enthalpy $I+P/\rho$: 
$$
   d(I+P/\rho)=C_p dT + \left ( \upsilon  - T \frac{\partial \upsilon}
   {\partial T} \right ) dP = C_p dT + 
   \upsilon \left ( 1 - \alpha T \right ) dP,
$$
we can derive the following equation for $PE_0$,
\begin{equation}
  \frac{d(PE_0)}{dt} = \int_{V_0} \rho_0 \left ( 
   C_{p0} \frac{DT_0}{Dt}  +  \left ( 1 - \alpha_0 T_0 \right )
    \frac{DP_0}{Dt} \right ) dV_0
   = \frac{dT_0}{dt} \int_{V_0} \rho_0 C_{p0} dV_0,
\end{equation}
which naturally provides the following equation for $T_0$:
\begin{equation}
      \frac{dT_0}{dt} = \frac{D(KE)+D(APE)+\dot{Q}-G(APE)}
      {\int_{V_0} \rho_0 C_{p0} dV_0} .
\end{equation}
We can now derive an evolution equation for $GPE_0$, using the relation:
\begin{equation}
     \frac{d(GPE_0)}{dt} = \int_{V_0} P_0' \frac{D\upsilon_0}{Dt} \rho_0 dV_0
\end{equation}
where $P_0'=P_0-P_a$. Now, expressing $d\upsilon= \upsilon \alpha dT
+ \upsilon \gamma dP$, where $\gamma$ is the isothermal expansion coefficient,
we arrive at the following expression:
\begin{equation}
   \frac{d(GPE_0)}{dt} = \int_{V_0} \rho_0 P_0' \left [
      \upsilon_0 \frac{DT_0}{Dt} - \upsilon_0 \gamma_0 \frac{DP_0}{Dt}
    \right ] dV_0 = \frac{dT_0}{dt} \int_{V} \alpha_0 P_0' dV_0
    {\rho_0 C_{p0}} dV_0
\end{equation}
noting again that the term proportional to $DP_0/Dt$ must vanish
from the arguments developed above, so that we simply have:
\begin{equation}
    \frac{d(GPE_0)}{dt} = \Upsilon_0 \left [
   D(KE) + D(APE) + \dot{Q}_{net} - G(APE) \right ]
\end{equation}
where
\begin{equation}
      \Upsilon_0 = \frac{\int_{V_0} P_0' \alpha_0 dV_0}
      {\int_{V_0} \rho_0 C_{p0} dV_0} .
\end{equation}
As a result, it follows that:
$$
   \frac{d(IE_0+P_a V_{ol,0})}{dt} = \frac{d(PE_0-GPE_0)}{dt}
$$
\begin{equation}
   = \left ( 1- \Upsilon_0 \right ) \left [ D(KE) + D(APE) 
  + \dot{Q}_{net} - G(APE) \right ]
\end{equation}
Let us now define the exergy part of the $IE_r + P_a V_{ol,r}$ as
\begin{equation}
      IE_{exergy} = IE_r - IE_0 + P_a (V_{ol,r}-V_{ol,0})
\end{equation}
The equation is:
$$
     \frac{d(IE_{exergy})}{dt} = 
     - W_r + \Upsilon_0 \left [ D(KE)+D(APE) + \dot{Q}_{net} - G(APE) 
    \right ]
$$
$$
    = (\Upsilon_0 - \Upsilon_{r,ke})D(KE) 
    + (\Upsilon_0 - \Upsilon_{r,ape})D(APE)
    + \Upsilon_0 \left [ \dot{Q}_{net} - G(APE) \right ]
$$
\begin{equation}
    - W_{r,mixing} + W_{r,forcing} .
\end{equation}
Again, the neglect of the terms proportional to $\alpha P/(\rho C_p)$
yields the following simplification:
\begin{equation}
    \frac{d(IE_0+P_a V_{ol,0})}{dt} \approx D(APE)+D(KE)+
  \dot{Q}_{net} - G(APE) ,
\end{equation}
\begin{equation}
    \frac{d(IE_{exergy})}{dt} \approx W_{r,mixing} - W_{r,forcing} .
\end{equation}
The corresponding energy flowchart is this time illustrated in
Fig. \ref{basic_energetics} (Panel III). This figure shows that when
$IE_r$ is decomposed into its dead and exergy part, a decoupling 
between the $KE/APE/IE_0$ and $IE_r-IE_0/GPE_r$ reservoirs appears
at leading order. Note, however, that the rates between the reservoirs
remain coupled, owing to the correlation between $D(APE)$ and
$W_{r,mixing}$, as well as between $G(APE)$ and $W_{r,forcing}$ discussed
in this paper, and which is a central topic of turbulent mixing theory.

\subsection{Separate evolution of $APE$ into $GPE$ and $IE$ components}

We conclude the evolution equation of energetics by further splitting
the $APE$ reservoir into its $GPE$ and $IE$ components. Using the
previous relations, one easily shows that: 
\begin{equation}
   \frac{d(AGPE)}{dt} = W - W_r \approx W - W_{r,mixing} + W_{r,forcing}
\end{equation}
$$
   \frac{d(AIE)}{dt} \approx W - B + G(APE) - D(APE) 
   - \left [ W - W_{r,mixing} + W_{r,forcing} \right ]
$$
\begin{equation}
   \approx - B + W_{r,mixing} - D(APE) + G(APE) - W_{r,forcing} .
\end{equation}
For seawater, it is generally found that $AIE$ accounts for around 
$10\%$ of the total $APE$, so that to a good approximation $APE\approx
AGPE$, which is implicit in the Boussinesq approximation. Equating
$d(AGPE)/dt$ with $d(APE)/dt$ amounts to requiring that $d(AIE)/dt \approx 0$.
By imposing that the forcing and mixing terms vanish separately,
one obtains:
\begin{equation}
         D(APE) \approx W_{r,mixing} - B ,
\end{equation}
\begin{equation}
         G(APE) \approx W_{r,forcing} ,
\end{equation}
which are equivalent to those of the L-Boussinesq and NL-Boussinesq model.
The corresponding energy flowchart is depicted in Fig. \ref{basic_energetics}
(Panel IV). The key feature of this figure is to reveal that the conversion
rates between $AGPE$ and $AIE$ are identical to those taking place between
$IE_r-IE_0$ and $GPE_r$, which appears to be where the coupling between
stirring and mixing fundamentally occurs.

\bibliographystyle{jfm}

\end{document}